# Title: Quantification of Nuclear Coordinate Activation on Polaritonic Potential Energy Surfaces

Authors: Shahzad Alam[1], Yicheng Liu[2], Russell J. Holmes[2], Renee R. Frontiera[1]*

Affiliations: [1]Department of Chemistry, University of Minnesota, Minneapolis, MN-55455

[2]Department of Chemical Engineering and Material Science, University of Minnesota, Minneapolis, MN-55455

Corresponding author: rrf@umn.edu

**Abstract:**

Polaritonic states, which arise from strong coupling between light and matter, show great promise in modifying chemical reactivity. However, reproducible enhancement of chemical reactions with polaritons is challenging due to a lack of understanding on how to launch wavepackets along productive reactive coordinates while avoiding unproductive local minima in the multidimensional potential energy landscape. Here we employ resonance Raman intensity analysis to quantify mode-specific nuclear displacement values in pentacene thin films and pentacene exciton-polaritons. We find that coupling significantly changes the potential energy landscape, including both enhancement and suppression of nuclear displacements. We demonstrate that controlling cavity parameters enables selective steering of vibronic wavepackets. Our approach provides a quantitative methodology for screening polaritonic catalysts and opens new avenues for designing reproducible and effective cavity-controlled chemistry.

**Main Text:**

Precise control of chemical reactivity is of paramount importance and is conventionally achieved by varying external conditions such as temperature and pressure, adding a catalyst, or synthetically modifying molecular structures. Recently, a number of works have shown that strong light-matter coupling can modify chemical reaction dynamics, attracting a surge of interest among researchers to manipulate material properties and steer chemical reactivity using light (*1–6*). Formation of these strongly coupled or polaritonic states offers an unconventional method to modify potential energy landscapes and thus control the rates, yields, and outcomes of a given reaction. In one notable experiment, Ebbesen *et al.* demonstrated that the rate of photoisomerization of spiropyran to merocyanine can be altered under strong coupling of light with molecular electronic states (*7, 8*). Furthermore, strong coupling has been successfully applied in achieving long-range energy (*9–11*) and charge transport (*12–14*), low threshold polariton lasers (*15–17*), and modified photophysical properties (*18–27*). Nevertheless, it's noteworthy that in many but not all cases, the rate or yield of chemical reactions seems to deteriorate within optical cavities, which has hindered progress in rational design approaches of polaritonic chemistry (*18, 28, 29*).

Precise reactivity control with light poses a significant design challenge primarily because of the complex nature of multidimensional potential energy surfaces (PESes). These involve a large number of vibrational degrees of freedom, described by the formula 3N-6 for a non-linear molecule, where N represents the number of atoms involved in the molecular system. Interactions between these nuclear degrees of freedom dictate how the wavepacket moves on the PES following photoexcitation. For example, when a specific nuclear coordinate is highly displaced in an excited state relative to the ground state, wavepackets generated after photoexcitation will be effectively launched along this coordinate, and the atoms will move rapidly. When those nuclear motions are involved in the reaction coordinate, this rapid motion can improve reactivity. To control photochemical reactions via electronic polaritons it is crucial to tailor the PES within the short-lived Franck-Condon region (*30–33*), since as the system relaxes on the excited state potential, it will rapidly leave the resonance conditions necessary for polariton formation. As such, the initial slopes of the excited state potentials in the photoexcitation region are crucial in launching wavepackets along certain reaction coordinates, and likely play a significant role in the downstream productivity of the photochemical reaction (*34, 35*). Nuclear coordinates with large initial slopes, or those with highly displaced potentials, will play the greatest role in directing the initial wavepacket motion. While numerous theoretical works have established the influence of strong-coupling induced modification of excitonic potential energy surfaces in the Franck-Condon region, direct experimental probes are challenging due to the short timescales involved (*31, 36, 37*).

There are many experimental degrees of freedom that can be varied to influence reactivity with polaritonic states. For instance, the strength of strong coupling, which is typically quantified by the branch splitting value, or the degree to which hybridization between light and matter modifies energy levels, is thought to be an important factor (*1*). The degree of strong light-matter coupling can be tuned by adjusting cavity parameters such as cavity detuning, pathlength, Q-factor, and refractive index as well as by adjusting molecular layer properties including active layer optical density, molecular orientation, and crystallinity. Often these parameters are interdependent. As such, it is challenging to rationally design cavities for modification of specific chemical reactions, as it is currently unclear how to best narrow this vast parameter space. To address this urgent need, a methodology is needed to systematically investigate the effects of these parameters on the polaritonic PES to identify those that are most impactful and can be effectively manipulated to achieve desired outcomes.

To experimentally investigate the effects of polariton formation on the multidimensional PES in the Franck-Condon region, it is necessary to have a technique capable of probing the ultrafast nuclear dynamics occurring on initial photoexcitation. Multiple experimental studies have tried to unravel the ultrafast dynamics of excitonic-polaritonic cavities through techniques such as transient absorption (TA) (*38, 39*) and time-resolved photoluminescence (PL) (*40–42*). In these TA and time-resolved PL measurements, researchers were able to assign the excited state spectral features to the polaritonic states, determine the excited state lifetime of polaritonic states, and deduce the reaction pathway involved. However, with these time-domain approaches, it can be challenging to determine the very earliest dynamics following photoexcitation, primarily due to coherent artifacts that overwhelm the optical response of the system around time zero (*43*). As such, many femtosecond techniques are limited in the ability to probe the very earliest timescale nuclear dynamics.

One well-established method which can probe and quantify changes in multidimensional potential energy surfaces in the Franck-Condon region is resonance Raman intensity analysis (RRIA). This steady-state technique involves measuring and fitting the absorption spectra and Raman excitation profiles (REPs), which quantify the dependence of the Raman cross-section for a particular vibrational mode as a function of the excitation energy. The experimental absorption spectra and REPs are modeled with a time-dependent wavepacket model (*44*). This analysis provides information about the excited state parameters, including the displacement ($\Delta$), a dimensionless value which describes the mode-specific nuclear displacement of a harmonic potential in a resonant excited state relative to the ground state. The $\Delta$ value is directly proportional to the slope of the excited state PES along a specific normal mode (*45*), meaning that modes with large $\Delta$ values will be efficient at launching excited state wavepackets along specific trajectories. This is illustrated in Fig. 1(a), where a higher $\Delta$ value for the polaritonic case indicates a steeper slope for the PES as compared to the molecular case. As a result, wavepackets will be more effectively launched under the polaritonic condition for that particular reaction coordinate, as compared to the uncoupled system. These dimensionless displacement $\Delta$ values are related to the mode-specific Huang-Rhys factors through the relationship $S = \Delta^2/2$. Huang-Rhys factors describe the strength of electron-vibration or electron-phonon coupling. A detailed discussion on RRIA has been provided in the supplementary material 1.1.

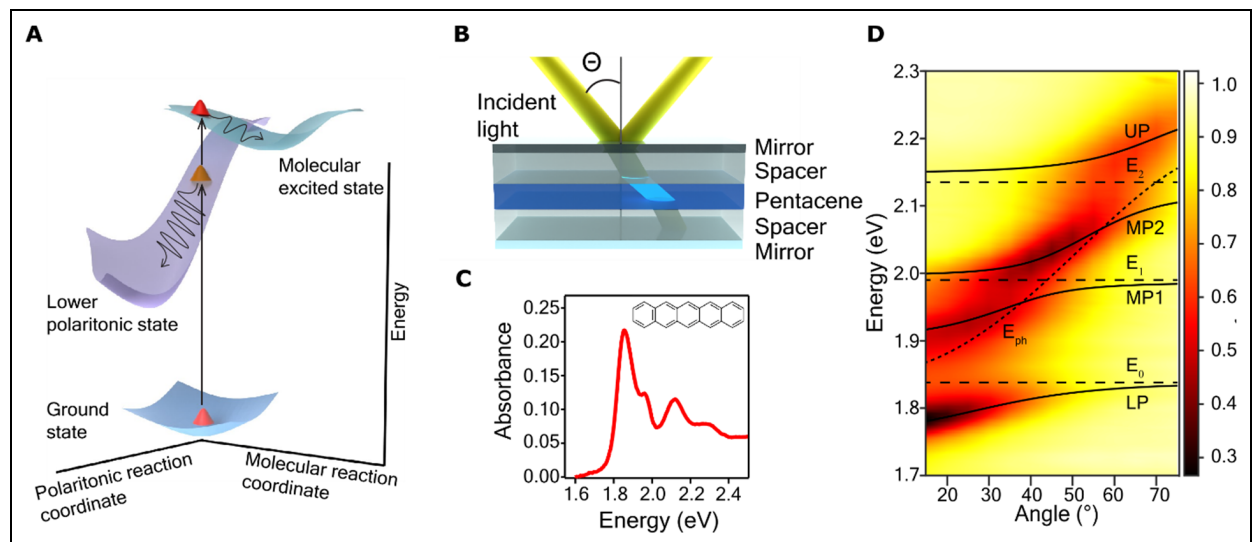

**Fig. 1. Characterization of pentacene excitonic polaritons.** (**A**) Potential energy diagram illustrating how polariton formation can modify reaction coordinates. (**B**) Schematic of the cavity structure, featuring a pentacene layer (27 or 39 nm), a tris(carbazol-9-yl)triphenylamine (TCTA) spacer layer, and semi-transparent 20 nm top and 70 nm bottom silver layers. (**C**) Absorbance spectrum of a 27 nm thick film of pentacene with its molecular structure measured at 0° incident angle. (**D**) Angle-resolved reflectivity map of the 57 nm TCTA/ 27 nm pentacene/ 57 nm TCTA cavity for s-polarized light illumination from 15° to 75° measured every 5°. The dashed black lines show the bare photon energy for this cavity ($E_{ph}$) and uncoupled excitonic transitions for pentacene ($E_0$, $E_1$, and $E_2$). The solid black lines are fitted dispersions via the coupled-oscillator model, based on the positions of reflectivity minima of each branch. LP, MP1, MP2, and UP stand for lower polariton, middle polaritons 1 and 2, and upper polariton, respectively. $E_0$, $E_1$, and $E_2$ are energies of the vibronic peaks of the uncoupled pentacene thin film, and $E_{ph}$ is the energy of the photon.

Resonance Raman intensity analysis has been applied previously to a number of molecular and materials systems, including initial excited-state torsional dynamics of cis-stilbene (*46*), electron-phonon coupling in quantum dots (*47, 48*), excited-state dynamics of MOFs (*49*), and Franck-Condon effects in electron transfer reactions (*50, 51*), among many others. A key advantage of this technique is that it readily obtains quantitative information regarding the shape of the excited state potential energy landscape in the initially-prepared photoexcitation region. It provides displacement values for all measured Raman-active modes, thus mapping out multiple dimensions of the PES. The instrumentation is relatively straight-forward as compared to time-domain approaches (*52*), and modeling code is freely available (*53*). It is important to note that the method relies on harmonic potential approximations, and that rapid vibronic dephasing can result in low magnitude signals. Additionally, the measurements are ensemble-averaged over all resonant species, which in the case of polaritonic systems can include both coupled and uncoupled molecular species. To date, resonance Raman intensity analysis has not been performed on polaritonic systems, although multiple Raman approaches for obtaining information about polaritonic potential energy landscapes have been proposed (*54, 55*) and implemented (*56, 57*). Here, by obtaining mode-specific displacements for a variety of polariton conditions, we are able to methodically screen how coupling parameters impact molecular PESs. This is important because it enables a quantitative methodology for determining how strong coupling impacts multidimensional potential energy landscapes and provides design guidelines for reaction-specific rational cavity design.

Here, we probed strongly coupled cavities containing either 27 nm or 39 nm thick films of pentacene, sandwiched between two silver mirrors and a spacer layer such that the total thickness of both the cavities is 140 nm (Fig. 1B). The spacer layer thickness is adjusted to center the pentacene thin film at the anti-node of the cavity. We compare these polaritonic systems to an uncoupled system consisting of a 27 nm thick pentacene film on a 70 nm silver mirror, so as to compare systems with similar molecular orientations. Detailed fabrication procedures are provided in the supplementary material 1.2. Fig. 1C presents the absorbance spectrum of a pentacene thin film. The primary singlet transition from $S_0$ to $S_1$ shows two features due to a Davydov splitting, with one at 1.85 eV ($E_0$) and the other at 1.99 eV($E_1$), along with a third feature at 2.14 eV which is the second singlet electronic state ($E_2$) (*58*). Fig. 1D presents the color plot of angle-resolved reflectivity dispersion for the 27 nm pentacene cavity, measured with s-polarized light. The $E_0$ peak absorption is polarization dependent and therefore the Raman and reflectivity measurements for pentacene as a thin film and in a cavity were carried out with s-polarized light for proper comparison. By fitting angle-resolved reflectivity spectra, an energy-angle dispersion can be constructed. Four distinct branches are observed in the dispersion for the 27 nm cavity, showing the anti-crossing characteristic of formation of four exciton-polariton states resulting from hybridization via their common interaction with the photon (*59, 60*). Here, LP, MP1, MP2, and UP denote the lower, first middle, second middle, and upper polaritonic branches, respectively. Similar behavior was observed for the 39 nm cavity and the data are shown in Fig. S1A. The dispersion relation in Fig. 1C was fit using a four-body coupled-oscillator model (*61*):

$$\begin{bmatrix} E_{ph}(\theta) & V_1 & V_2 & V_3 \\ V_1 & E_{ex0-0} & 0 & 0 \\ V_2 & 0 & E_{ex0-1} & 0 \\ V_3 & 0 & 0 & E_{ex0-2} \end{bmatrix} \begin{bmatrix} \alpha \\ \beta \\ \gamma \\ \sigma \end{bmatrix} = \varepsilon \begin{bmatrix} \alpha \\ \beta \\ \gamma \\ \sigma \end{bmatrix}$$

where $V_1, V_2, V_3$ are interaction potentials, $\{|\alpha|^2, |\beta|^2, |\gamma|^2, |\sigma|^2\}$ are the Hopfield coefficients of the photonic and excitonic components, and $\varepsilon$ is the energy eigenvalue. The Hopfield coefficients were also extracted for the two cavities for LP and MP1 and are shown in Fig. S2. The cavity photon $E_{ph}(\theta)$ is modeled as a Fabry-Pérot mode as:

$$E_{ph}(\theta) = E_{ph}(0)\left(1 - \left(\frac{\sin(\theta)}{n}\right)^2\right)^{-1/2}$$

where $E_{ph}(0)$ is the cavity photon energy at normal incidence and $n$ is the effective refractive index. The dispersion for both cavities was modeled numerically with $E_0$, n, and interaction potentials as the fitting parameters, presented in Table S1. The branch splitting values, defined here as the energy gap between the lower polariton (LP) and middle polariton 1 (MP1), are 143 meV and 168 meV for the 27 nm pentacene and the 39 nm pentacene cavity, respectively.

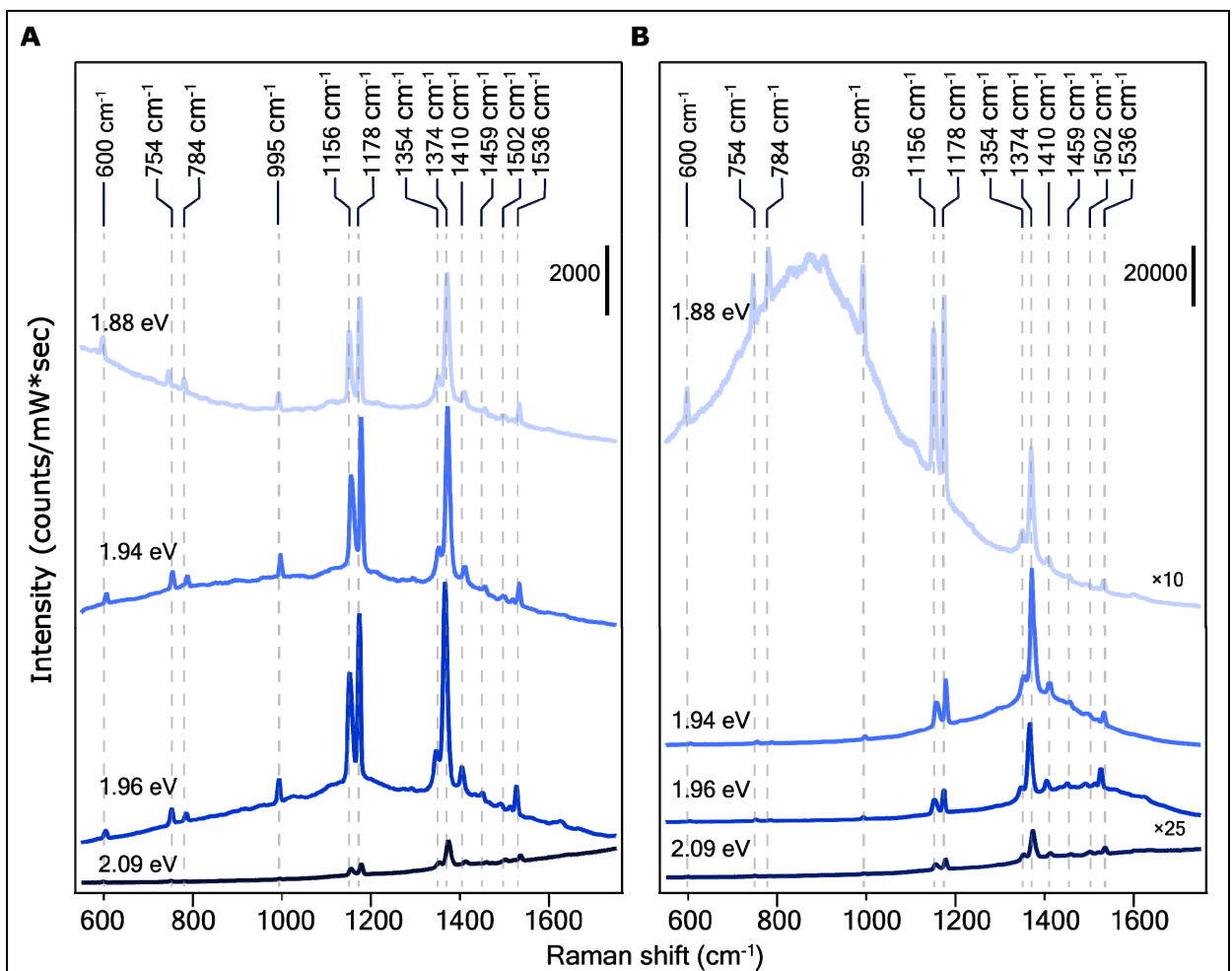

**Fig. 2. Resonance Raman spectroscopy.** Resonance Raman spectra of a 27 thin film of pentacene on a silver mirror (**A**), and a strongly coupled optical cavity containing 27 nm of pentacene (**B**), measured with 2.09, 1.96, 1.94 and 1.88 eV excitation sources.

We used angle-resolved resonance Raman spectroscopy to examine the 27 nm and 39 nm pentacene cavities. Fig. 2 displays the resonance Raman spectra at indicated excitation energies for the pentacene thin film on a silver mirror (A) and the 27 nm pentacene cavity (B) measured at 15° (also see supplementary material 1.3). The resonance Raman signal comes from both coupled and uncoupled pentacene molecules since the excitation is resonant with both the singlet and polariton absorption bands. However, the suppressed Raman intensity of all vibrational modes with 2.09 eV excitation for the pentacene cavity as compared to the thin film suggests a significant impact from polaritonic states. The resonance Raman spectra for the cavity show an enhancement in the Raman intensity with 1.88, 1.94 and 1.96 eV excitation, which we attribute in part to the Purcell effect, or the enhancement of spontaneous Raman scattering through coupling to the enhanced zero point radiation field in the cavity, as discussed below (*62*). In contrast, the thin film only exhibits enhancement due to the resonance effect, with a minor enhancement from the underlying Ag mirror. The signal magnitude is approximately one order of magnitude smaller as compared to the resonant cavity. The resonance Raman spectra also contain a photoluminescence (PL) background for both the thin film and cavity. The PL signal in the cavity exhibits angle dependence (Fig. S34), indicating an origin from the LP state, while the PL from the thin film arises from the $S_1$ state (Fig. S35), remaining isoenergetic across angles. The resonance Raman spectra for the two cavities with all four excitation energies are shown in Figures S32 and S33.

The Raman spectrum of pentacene has several intense peaks, including those at $\nu_5$ (1156 cm$^{-1}$, CH sides in-plane bending), $\nu_6$ (1178 cm$^{-1}$, peripheral CH in-plane bending), and $\nu_8$ (1374 cm$^{-1}$, CC aromatic stretch). Other Raman active modes include $\nu_1$ (600 cm$^{-1}$, ring twisting), $\nu_2$ (754 cm$^{-1}$, out of plane CH stretch), $\nu_3$ (784 cm$^{-1}$, ring breathing), $\nu_4$ (995 cm$^{-1}$, in plane H wagging of terminal rings), $\nu_7$ (1354 cm$^{-1}$, CC stretch and in-plane CH bending), $\nu_9$ (1410 cm$^{-}$1, CC backbone symmetric stretch and in-plane CH bond deformation), $\nu_{10}$ (1459 cm$^{-1}$, CC anti-symmetric stretch), $\nu_{11}$ (1502 cm$^{-}$1, CC stretch and in-plane CH bending), and $\nu_{12}$ (1536 cm$^{-1}$, CC backbone symmetric stretch and in-plane CH bond deformation) (*63*, *64*). Most of the Raman peaks show resonance enhancement in the pentacene thin film and the cavity. However, the $\nu_8$ mode for pentacene cavity is enhanced to a greater extent than any other mode relative to the thin film. In contrast, the lower frequency modes (below 1000 cm$^{-1}$) exhibit less enhancement in the pentacene cavity. In general, the relative ratio of Raman mode amplitudes with different excitation sources remains similar for the pentacene thin film, whereas significant differences are observed for the pentacene cavity. For instance, Fig. 2B shows that at excitation energies of 1.88 eV, 1.94 eV, 1.96 eV, and 2.09 eV, the vibrational mode amplitude ratio between $\nu_8$ and $\nu_6$ modes are 0.5, 2.8, 3, and 2.6 inside the cavity, respectively. In contrast, outside the cavity, the corresponding ratios are 1.1, 1.2, 1.2, and 2.3. These differences indicate that strong coupling does in fact induce mode-specific changes in the excited state molecular geometry within the cavity, and that polariton formation does modify excitonic potential energy surfaces.

To quantitatively investigate the impact of light-matter coupling on displacement ($\Delta$), we conducted resonance Raman intensity analysis on the pentacene thin film and both cavities. This involves an iterative global fit of both the absorption spectra and the Raman excitation profiles to obtain best fit values for the dimensionless displacements, or $\Delta$ values, for each Raman active coordinate (*53*). The advantage of this time-dependent wavepacket approach is that we obtain quantitative metrics as to the molecular distortion along each Raman active coordinate, or in other words, a clear map of which normal modes are activated upon photoexcitation. One drawback of this analysis is that we do not obtain the sign of the displacement, and thus can obtain only the magnitude of the nuclear change and are insensitive to the phase of the vibrational motion. Another limitation of applying RRIA approach to polaritonic systems is that it is historically applied to molecular-based systems (*44*). Given the hybrid nature of polaritons, the assumptions inherent in this technique might not fully capture the complexities introduced by strong light-matter coupling. However, for the purposes of our analysis, we assume that these limitations would affect all modes uniformly or in a predictable manner, and we can extract relative changes in mode-specific displacements. Therefore, our relative comparisons of displacement values ($\Delta$ values) remain valuable for understanding the overall excited-state vibrational dynamics within the system.

In our analysis, we considered the enhancement of Raman scattering in the cavity due to three factors: (a) amplification of the applied electric field within the cavity, (b) the Purcell effect, or enhancement of the spontaneous Raman scattering rate by the increased zero-point radiation field, and (c) Raman scattering collection efficiency as impacted by cavity effects on the Raman scattering spatial distribution (*65–67*). Proper normalization of the Raman intensity with these three enables accurate quantification of the impact of polariton formation on the molecular potential energy surface. Details of these enhancement normalization procedures are provided in the supplementary material section 1.4, and experimental Raman and absorption cross-sections are presented in supplementary material 1.5. For the pentacene thin film on the silver mirror, we calculated the enhancement factor associated with the mirror's effect on Raman scattering and subsequently accounted for this effect. Fig. 3A and C display the calculated and experimental absorption spectra for a pentacene thin film and the 27 nm cavity using the parameters detailed in Table S2-S4. For the pentacene thin film, the absorption band exhibits multiple features resulting from Davydov splitting. Therefore, in order to achieve a good agreement between calculated and experimental spectra a highly displaced mode was added at 730 cm$^{-1}$, which accounts for all modes not included in the calculations (*68, 69*). We also conducted RRIA analysis on the pentacene thin film by excluding the additional frequencies. The resulting modeled absorption and REPs are shown in Figures S29 and S30. While the absorption spectrum is not accurately modeled in this case, the Δ values were found to be similar.

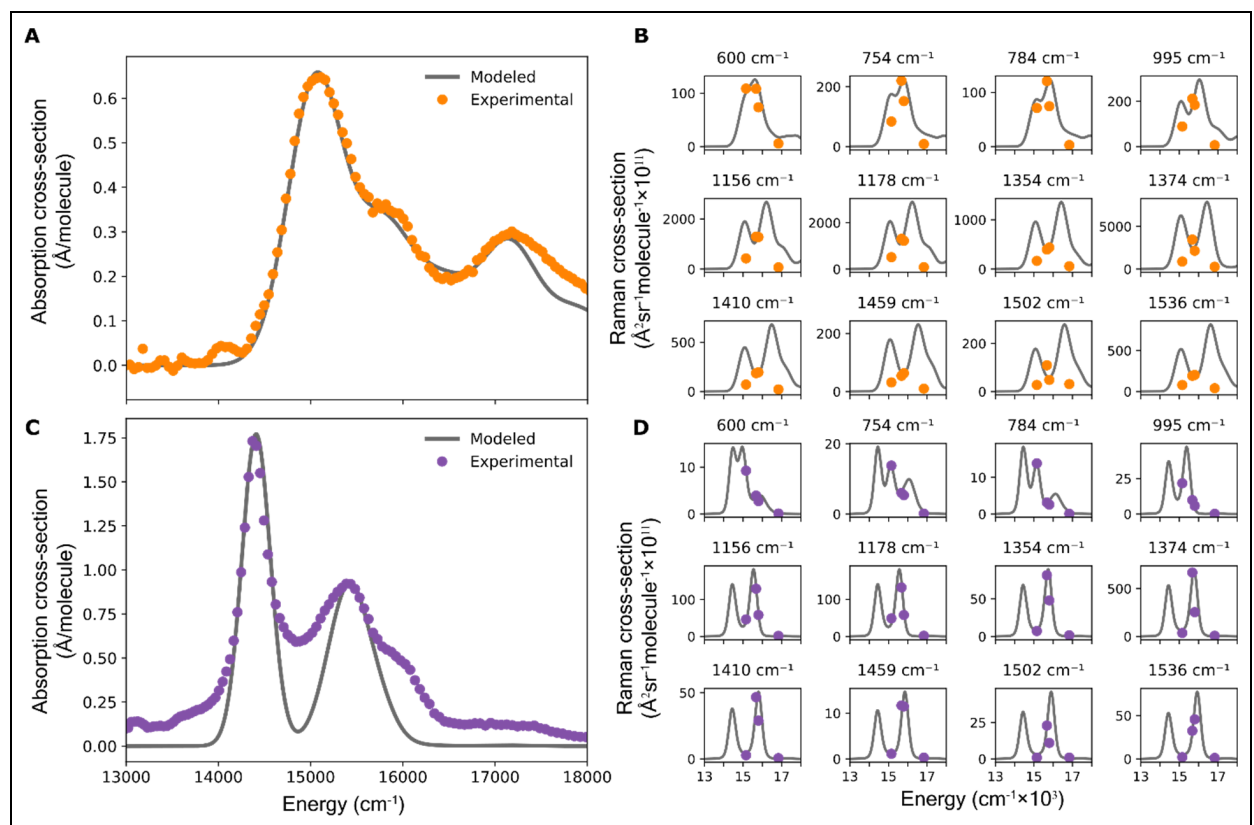

**Fig. 3. Raman excitation profiles and global modeling.** Experimental and calculated absorption spectra of pentacene thin film on a silver mirror (**A**) and 27 nm cavity (**C**). Raman excitation profile for pentacene thin film (**B**) and 27 nm cavity (**D**) at 15°.

The angle resolved reflection spectra for the pentacene cavity display four polaritonic branches, and we focus our analysis on the lower polaritonic states. We use a two-state model, considering resonant contributions from MP1 and LP, although the delta values for MP1 are generally quite small (supplementary material 2 and Fig. S4), indicating very rapid decay from this state. This model provides good agreement between the experimental and calculated absorption spectra and REPs for the pentacene cavity. The REPs for the pentacene thin film and the 27 nm pentacene cavity are shown in Fig. 3B and D as measured at 15 degrees respectively. We show modeled REPs and the absorption spectra for both cavities at all angles and pentacene thin film at 30° and 45° in Figures S5-S28, in which the orange and purple points represent experimental data, and grey lines present the global model.

Following the fitting of experimental and calculated absorption and REP's, $\Delta$ values were extracted for all Raman active vibrational modes, as presented in Fig. 4A. In this figure, the $\Delta$ values are plotted against the photonic weight of the LP state on the x-axis and the branch splitting between the LP and MP1 states on the y-axis. Each data point represents the $\Delta$ value for a specific Raman active mode in a particular cavity configuration, with color indicating the magnitude of the $\Delta$ value. Circles and triangles correspond to cavity configurations with 27 nm and 39 nm pentacene cavity, respectively, while rectangular blocks represent the $\Delta$ values for the thin film, provided for comparison.

Excitingly, we see that the mode-specific displacements change with different coupling conditions, indicating that cavity tuning can activate specific nuclear coordinates. For the uncoupled thin film, the modes $v_8$ ($\Delta = 0.35$), $v_6$ ($\Delta = 0.22$) and $v_5$ ($\Delta = 0.22$) modes exhibit the largest $\Delta$ values at 15°. These normal modes undergo the most rapid change in the geometry on the excited state PES upon photoexcitation. If polariton formation alters the slope of the PES in the Franck-Condon region along a normal coordinate, then the corresponding $\Delta$ values for the cavity are expected to be different than the thin film. Interestingly, for the 27 nm cavity (measured at a 15°, ~0.47 photonic weight), the corresponding $\Delta$ values for these modes are 0.28, 0.14, and 0.14, respectively. When the $\Delta$ values are lower for the polaritonic state as compared to the uncoupled state, this means that these coordinates are less distorted upon excitation in the cavity. For rest of the normal modes $v_1$, $v_2$, $v_3$, $v_4$, $v_7$, $v_9$, $v_{10}$, and $v_{11}$ and $v_{12}$, the pentacene thin film displays slightly larger or similar displacements (0.045, 0.072, 0.048, 0.075, 0.17, 0.12, 0.07, 0.07, and 0.13 respectively) compared to the 27 nm cavity (0.04, 0.049, 0.048, 0.07, 0.10, 0.075, 0.04, 0.07, and 0.09). For this particular polaritonic condition, the $\Delta$ values for all normal modes in the cavity are smaller than or similar to those of the thin film, indicating that the excited state PES is more similar to the ground state PES. However, as we will see in subsequent analysis, different coupling conditions can result in an excited-state PES that is significantly different from the ground-state PES, leading to a higher $\Delta$ value compared to the thin film. This means that strong coupling can influence $\Delta$ values in both directions, either increasing or decreasing nuclear coordinate changes relative to the ground state configuration.

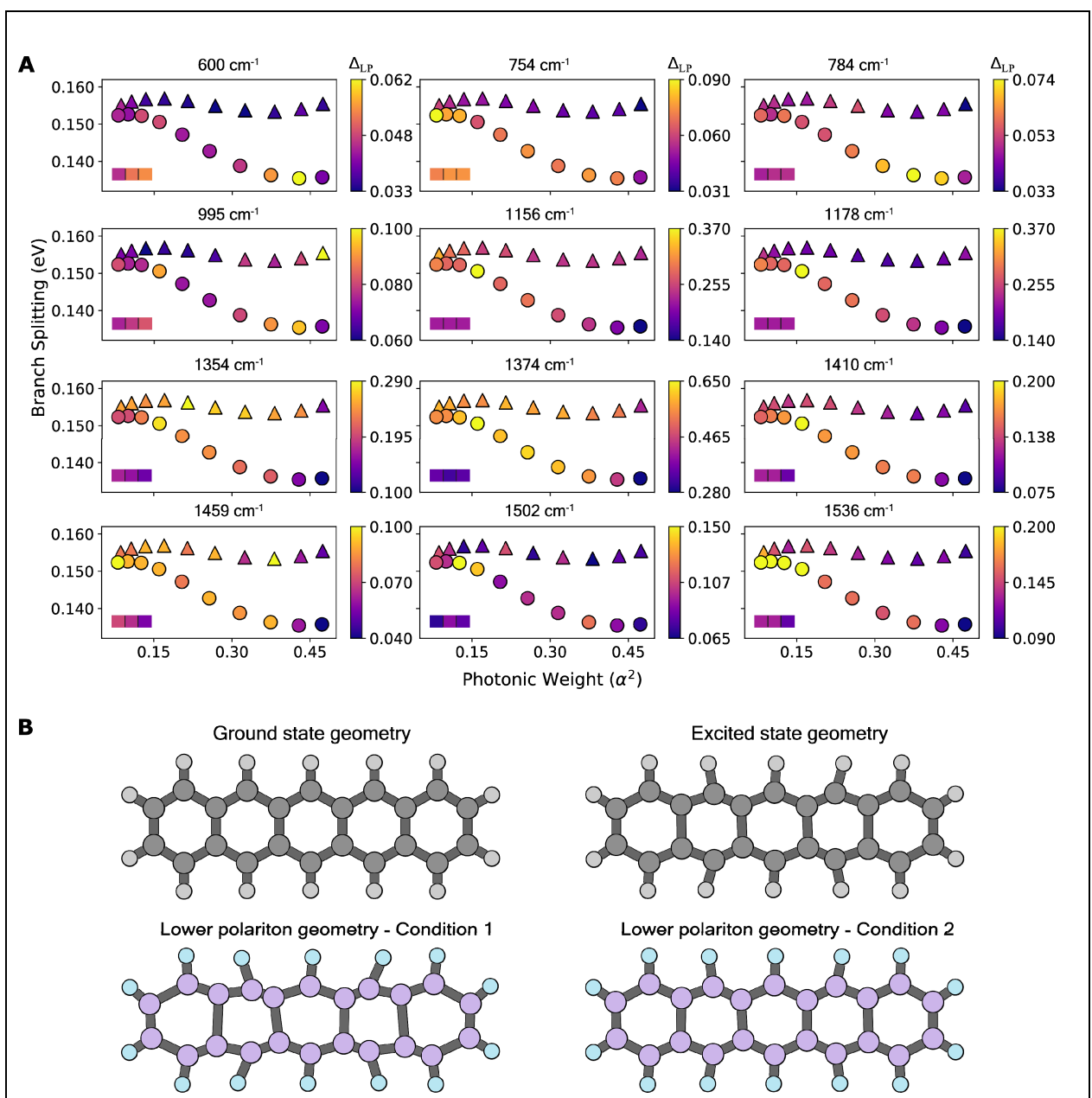

**Fig. 4. Polaritonic modification of potential energy landscapes.** (**A**) Displacement ($\Delta_{LP}$) for the lower polaritonic state for 27 nm (circle) and 39 nm (triangle) pentacene cavities plotted against the branch splitting and photonic weight. The color of the point indicates the displacement value. The pentacene thin film $\Delta$ values are indicated by the color of the rectangular blocks in the bottom left corner of each plot, where the first, second, and third blocks correspond to measurements at 15°, 30°, and 45°, respectively. (**B**) Comparison of equilibrated geometries for ground state, excited state, and lower polariton state (under different strong coupling conditions) conditions. Conditions 1 and 2 correspond to the 27 nm cavity with approximately 16% (45° measurement) and 47% photonic character (15° measurement), respectively. The $\Delta$ values were scaled by a factor of 2 to emphasize the distortion.

Another factor that impacts the $\Delta$ values is the strength of exciton-photon coupling as measured by the branch splitting. To quantify the impact of branch splitting on $\Delta$ values, we examined the 39 nm pentacene cavity, defined by a higher branch splitting with an energy gap of 168 meV between MP1 and LP. The $\Delta$ values for this cavity, depicted in Fig. 4A by triangles, are as follows: $\nu_1$ (0.035), $\nu_2$ (0.031), $\nu_3$ (0.033), $\nu_4$ (0.10), $\nu_5$ (0.23), $\nu_6$ (0.21), $\nu_7$ (0.15), $\nu_8$ (0.42), $\nu_9$ (0.095), $\nu_{10}$ (0.051), $\nu_{11}$ (0.075), and $\nu_{12}$ (0.10), measured at a 15° angle. Notably, only $\nu_8$ and $\nu_4$ modes have a higher $\Delta$ value and the wavepacket is launched more rapidly along these normal modes in the excited state PESs as compared to the pentacene thin film while the other modes are suppressed.

For lower frequency modes, defined here as those with values less than the branch splitting energy, the $\Delta$ values are smaller or similar under strong coupling conditions as compared to thin film. Conversely, for higher frequency modes, $\Delta$ values often exceed those of the thin film. This could be due to the fact that for low frequency vibrations, multiple quanta of vibrational excitation are required for the LP to be isoenergetic with the MP1 state. However, for higher vibrational frequencies, the MP1 and LP become nearly isoenergetic at low vibrational quantum number. Therefore, the interaction of the wavepacket in the LP state with MP1 along a high frequency coordinate distorts the molecular geometry to a greater extent and yields higher $\Delta$ values than in thin film. This may suggest that the MP1 state plays an influential role in modulating the $\Delta$ values of LP state.

Our results reveal that the extent of PES modification varies with the different cavity conditions, influenced by the photonic weight and the energy gap between the LP and MP1 states, leading to significant changes in $\Delta$ values as shown in Fig. 4. This suggests that specific modes can in fact be activated based on proper choice of cavity conditions, a key insight needed for rational design of polaritonic chemistry. For example, the $\Delta$ value for the most active mode, $\nu_8$, varied from 0.28 to 0.65 depending on the cavity configuration. When the photonic weight reached 16% and the energy gap was 150 meV, mode $\nu_8$ exhibited a $\Delta$ value of 0.65, indicating that the pentacene molecule undergoes its most significant shift in equilibrium geometry along this normal coordinate inside the cavity. Therefore, Fig. 4A showcases cavity design can specifically activate or deactivate a particular nuclear coordinate for a desired outcome.

One point to note is that the REPs arise from a global fitting process, and there can be other sets of parameters which provide good fits. Adjusting the $\Delta$ value impacts the modeled absorption spectra, particularly for Raman modes strongly coupled to the excited electronic state, such as $\nu_8$ (1374 cm$^{-1}$) which means that the $\Delta$ value range can be quite constrained. Also, the allowable range for modifying the $\Delta$ value of any given mode is small compared to the broader range of $\Delta$ values observed in our angle-resolved RRIA analysis. Some REPs, such as those for $\nu_1$, $\nu_2$, $\nu_3$, $\nu_4$, $\nu_5$, and $\nu_6$ for the 15° measurement, align well with the model, making them straightforward to fit. In contrast, for modes like $\nu_7$ and $\nu_8$, particularly at 15°, the experimental data points do not align perfectly with the modeled curve. In these cases, we aim to match the overall REP intensity to achieve the best possible fit. We follow a similar procedure to fit the REPs at all other angles to stay consistent. Fig. S31 further illustrates how adjusting $\Delta$ values affects the REPs for the 27 nm pentacene cavity at 15°, providing justification to the narrow range of parameters which provide a satisfactory global model. As such, the $\Delta$ values presented in Fig. 4a are representative of the modification to the PES caused by polariton formation, and the differences are meaningful.

Figure 4B shows pentacene structures under different molecular and polaritonic conditions, assuming harmonic potential and equilibrated geometries using the Δ values from RRIA. To depict the geometric distortions in excited state equilibrium geometry caused by polariton formation, we employed a numerical method (*70, 71*) to determine the Franck-Condon relaxed excited state structure using the following equation (*72*):

$$\delta_i = 5.8065 \sum_j A_{ji} \omega_j^{-1/2} \Delta_j$$

Here $\delta_i$ represents the structural change between the ground and excited states along internal coordinate i (measured in angstroms for bond lengths and degrees for angles); $A_{ji}$ is a matrix element connecting normal coordinate j to internal coordinate i; $\omega_j$ is the frequency of normal mode j, and $\Delta_j$ is the dimensionless displacement along normal coordinate j. This involves projecting dimensionless normal mode displacements onto cartesian coordinates, as illustrated in Fig. 4B. We assume that the displacements are all positive, which means that the method does not determine the relative phase of the impacted vibrational motions. Details about the numerical methods are provided in the supplementary material 1.6. Fig. 4B compares the ground state, excited state, and lower polariton state (under two different conditions) geometries, revealing the extent of geometric distortion upon polariton formation. Conditions 1 and 2 correspond to the 27 nm cavity with approximately 16% (at 45°) and 47% photonic character (at 15°), respectively. Condition 2, where the Δ values are smaller or comparable to those of the thin film, closely resembles the ground-state geometry. In contrast, condition 1 shows a geometry with more pronounced distortions than the molecular excited-state geometry, as indicated by its higher Δ values compared to the thin film. Under all conditions, the molecular distortions are primarily in the plane of the molecule, likely because these are the modes that most strongly couple to the electronic transition.

Overall, our findings conclusively demonstrate how polariton formation modifies the potential energy surfaces of a pentacene thin film embedded in a strongly coupled optical cavity. By carefully manipulating the cavity and molecular layer parameters, different mode-specific displacement (Δ) values can be achieved. We find that the potential energy surface along all Raman active modes of the pentacene cavity can be modified relative to a thin film of pentacene, with the $v_8$ mode showing the most pronounced modification. This work shows that by adjusting cavity parameters like photonic weight and energy gap, molecular behavior can be influenced leading to control over potential energy surfaces. In addition, these results suggest a promising avenue for selectively launching wavepackets on the excited state PES along a particular set of nuclear coordinates. While further investigation is needed to establish the link between activating or deactivating a specific set of modes and the downstream reaction outcome, our results provide a basis for cavity design. They also show that subtle changes in cavity preparation can lead to vastly different reaction coordinates. By understanding how polariton formation in different conditions influence the PES in the Franck-Condon region, strategies can be devised to steer wavepackets away from the local minima of the reactant side, potentially avoiding situations where polaritons worsen the chemical reactivity. This study offers a fundamental understanding of polariton-mediated modification to the PES and can be used in exploring the potential applications of cavity-controlled chemical reactions and devices.

**Acknowledgements:**

The work was supported by the National Science Foundation through University of Minnesota MRSEC under DMR-2011401 and the U.S. Department of Energy Grant SC-0023374 (R. R. F. and S. A.). R. J. H. acknowledges support from Ronald L. and Janet A. Christenson. Y. L. acknowledges support from a UMN College of Science and Engineering Fellowship.

**Data Availability:**

All data are freely available upon request, or through the Data Repository for the University of Minnesota (DRUM) at https://hdl.handle.net/11299/166578

**List of Supplementary Materials:**

Materials and Methods
Supplementary Text
Figs. S1 to S35
Tables S1 - S4
References (*1–11*)

# Supplementary Materials for

## Quantification of Nuclear Coordinate Activation on Polaritonic Potential Energy Surfaces

Shahzad Alam[1], Yicheng Liu[2], Russell J. Holmes[2], Renee R. Frontiera[1]*

*Corresponding author: rrf@umn.edu

**The PDF file includes:**

    Materials and Methods
    Supplementary Text
    Figs. S1 to S35
    Tables S1 to S4
    References

1. Materials and Methods

1.1 <u>Resonance Raman intensity analysis theory</u>

The absorption spectra and Raman excitation profile were modelled using Heller's time-dependent formalism, which utilizes wavepacket propagation on the excited state potential energy surface (*1, 2*). The absorption and Raman cross sections are described by equations (1) and (2):

$$\sigma_A = \frac{4\pi e^2 M^2 E_L}{6\hbar^2 c} \int_{-\infty}^{\infty} \langle i|i(t)\rangle \times exp\left[i(E_L + E_i)\frac{t}{\hbar} - \frac{\Gamma|t|}{\hbar}\right] dt \qquad (1)$$

$$\sigma_R = \frac{8\pi e^4 M^4 E_S^3 E_L}{9\hbar^6 c^4} \left|\int_0^{\infty} \langle f|i(t)\rangle \times exp\left[i(E_L + E_i)\frac{t}{\hbar} - \frac{\Gamma t}{\hbar}\right] dt\right|^2 \qquad (2)$$

In this expression, M represents the transition length, $E_L$ is the incident photon energy, $\Gamma$ is the homogenous linewidth, $|i\rangle$ is the initial ground state vibrational wavefunction of the ground electronic state, and $|i(t)\rangle = e^{-iHt/\hbar}|i\rangle$ is the time-evolved initial wavefunction under the excited state potential (*H*). Similarly, the Raman cross section is given by equation (2). Here, $E_s$ is the energy of the scattered photon, and $|f\rangle$ is the final excited vibrational level on the ground state surface, and the $\langle f|i(t)\rangle$, referred to as the Raman overlap, describes the overlap of the initial and final wavefunctions at time t.

In resonance Raman scattering, electronic excitation leads to a redistribution of charge, causing a displacement of normal coordinates and changes in the polarizability. Notably, Raman mode enhancement occurs only if the normal coordinate is significantly displaced upon photoexcitation. This can be intuitively understood in terms of equation (2). At t = 0, the system starts in the initial vibrational state $|i\rangle$ of the ground electronic state. Upon interaction with the incident photon, the wavepacket is propagated on the excited-state potential surface, influenced by the excited-state potential (*H*). If the ground and excited surfaces are not displaced to a greater extent, the propagated wavepacket $|i(t)\rangle$ never moves far enough to gain a good overlap with the final state $|f\rangle$, resulting in weak resonance enhancement. Therefore, only normal modes which are significantly displaced will be able to achieve a high Raman overlap value and will show strong resonance Raman enhancement.

The multidimensional Raman overlap is given by the product of $\langle f|i(t)\rangle$ for the Raman active mode and $\langle i|i(t)\rangle$ for all other modes, with the fundamental Raman scattering described by equations (3) and (4):

$$\langle 0|0(t)\rangle = exp\left[-s(1-e^{-i\omega t}) - \frac{i\omega t}{2} - \frac{iE_0 t}{\hbar}\right] \qquad (3)$$

$$\langle 1|0(t)\rangle = \pm s^{1/2}(e^{-i\omega t} - 1)\langle 0|0(t)\rangle \qquad (4)$$

where $s = \Delta^2/2$ and is also known as the Huang-Rhys factor and the $\Delta$ is the dimensionless nuclear displacement (*3*).

### 1.2 <u>Cavity fabrication and characterization</u>

Two different microcavities were fabricated for the purpose of this work, consisting of 57 nm 4,4',4-tris(carbazol-9-yl)triphenylamine (TCTA)/ 27 nm pentacene/ 56 nm TCTA (cavity 1) and 50 nm TCTA/ 39 nm pentacene/ 51 nm TCTA (cavity 2) sandwiched between bottom and top reflectors of 70 nm and 20 nm of Ag, respectively (Fig. 1B). The active absorber material in the microcavity is pentacene, whose molecular structure and absorbance are shown in Fig. 1C. TCTA was employed as the spacer layer to enable tuning of the pentacene thickness while keeping the cavity mode energy fixed. All of the layers were thermally evaporated at a base pressure of < 9×10⁻

[7] Torr on cleaned glass slides with a deposition rate of 0.2 nm/s at a substrate temperature of 25 °C. A quarter of each microcavity was covered by a shadow mask during the top mirror deposition to also enable measurements on thin films.

Angle-resolved reflectivity spectra were measured from 15° to 75° with a step size of 5° using a variable angle spectroscopic ellipsometer (VASE, J. A. Woollam). Experimental spectra were fit via multipeak fitting (Gaussian). A four-body coupled oscillator model was applied to fit the experimentally measured dispersion with coupling strength and cavity mode energy as fitting parameters. Table S1 summarizes all the fitting parameters for both the 27 nm and 39 nm pentacene cavities. Optical constants for pentacene were extracted from pentacene thin films deposited on a 70 nm Ag and a glass substrate. An isotropic Cauchy model was used to extract the optical constants via point-by-point fitting. To ensure the fitting is physically meaningful, a Kramers-Kronig test was performed to ensure the unconstrained fitting with Cauchy model obey the physical relationship between refractive index and extinction coefficients. A transfer matrix formalism was used to simulate the angle-resolved reflectivity spectra of pentacene cavities using the extracted optical constants.

**Table S1.** Coupled oscillator fitting parameters for the 27 and 39 nm pentacene cavities.

|  | $E_0$ (eV) | $V_1$ (meV) | $V_2$ (meV) | $V_3$ (meV) | $n$ |
|---|---|---|---|---|---|
| 27 nm cavity | 1.85 | 71.4 | 33.4 | 33.0 | 1.88 |
| 39 nm cavity | 1.85 | 84.0 | 43.1 | 62.4 | 1.78 |

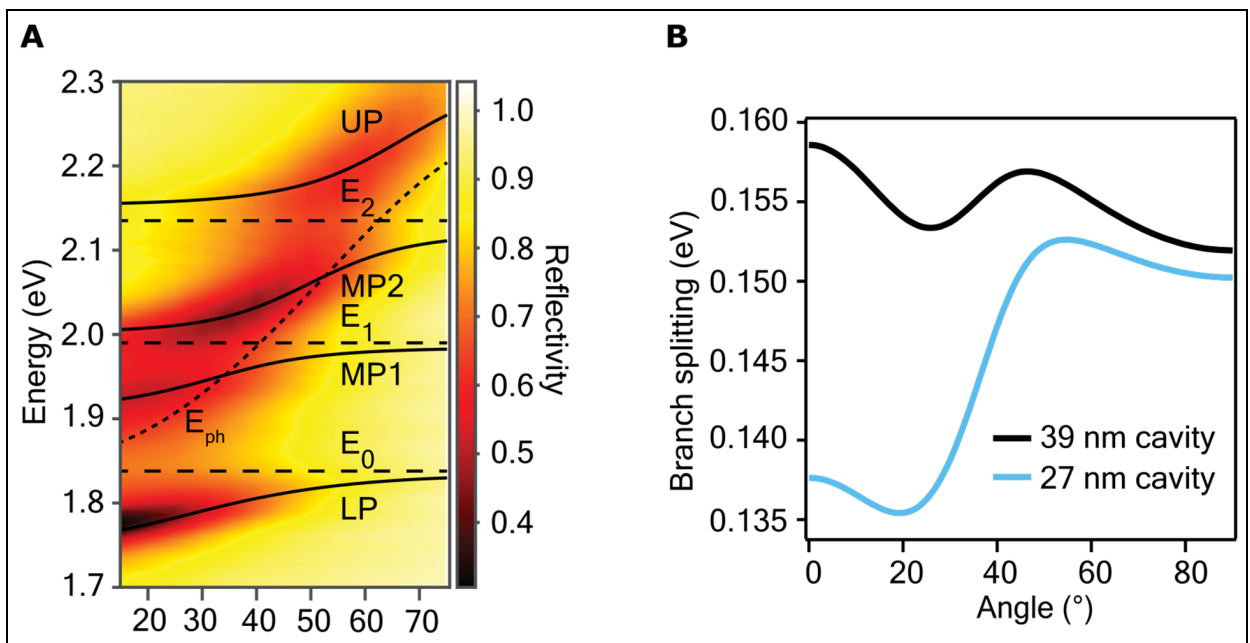

**Fig. S1. (A)** Angle-resolved reflectivity map of the 50 nm TCTA/ 39 nm pentacene/ 51 nm TCTA cavity for TE polarization from 15° to 75°. The dashed black lines show the bare photon energy for this cavity and uncoupled excitonic transitions for pentacene. The solid black lines are fitted dispersions via a coupled-oscillator model, based on the positions of reflectivity minima of each branch. **(B)** Energy gap between the LP and MP1 state for both the cavities.

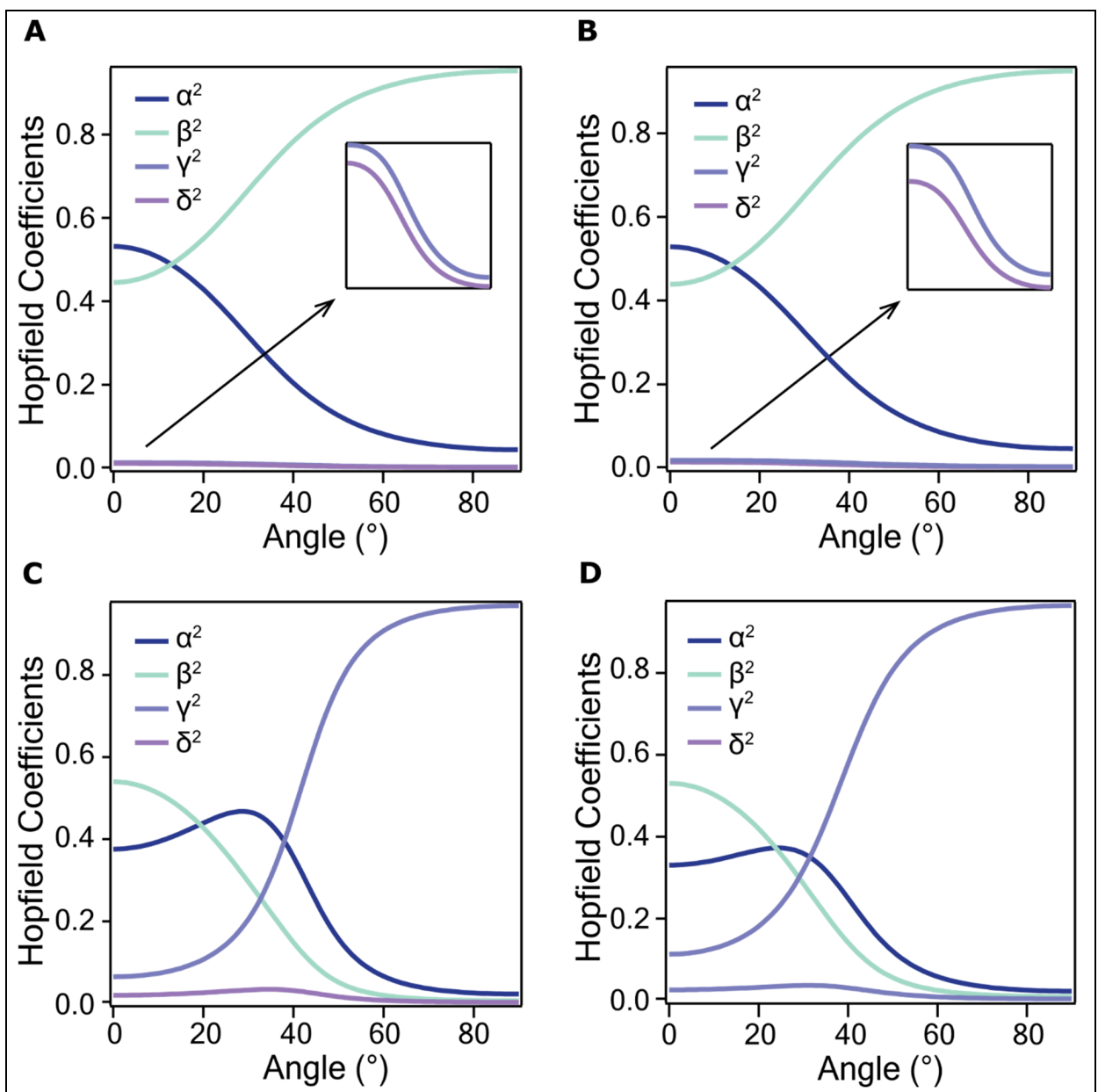

**Fig. S2.** Hopfield coefficients for the LP state are depicted for 27 nm and 39 nm pentacene cavity in **(A)** and **(B)**, and for the MP1 state in **(C)** and **(D)** respectively. Here, $\alpha^2$, $\beta^2$, $\gamma^2$, and $\delta^2$ coefficients represents the photonic, first vibronic peak of $S_1$ ($E_0$), second vibronic peak of $S_1$ ($E_1$), and $S_2$ ($E_2$), respectively.

## 1.3 Resonance Raman measurement

Resonance Raman spectra were acquired using excitation wavelengths of 2.09, 1.96, 1.94, and 1.88 eV. A 5 cm focal length lens was used to focus the excitation light onto the sample, placed on a goniometer for angle-resolved Raman spectra measurements. The scattered light was collected in a backscattering geometry and was directed into a spectrometer (Acton SP2500, PI) with a CCD detector (PIXIS 100BX, PI).

## 1.4 Enhancement factor calculation

The enhancement factor for Raman intensities due to (a) electric field amplification by the cavity, (b) Purcell effect, and (c) solid angle were calculated using the equation (5) (*7–9*):

$$Enhancement\ factor = \left((E/E_0)^2_{cavity}\right)\left(\frac{3}{4\pi^2}\frac{\lambda_s^3}{n^3}\frac{Q}{V}\right)\left(\frac{4\pi^2\omega_0^2}{\lambda_s^2}\right) \quad (5)$$

the first factor represents the electric field amplification inside the cavity, where $(E/E_0)_{cavity}$ and $(E/E_0)_{thin\ film}$ are the normalized electric fields of pentacene cavity and thin film with respect to the incident electric film at a specific excitation wavelength. Transfer matrix simulations were applied to calculate the electric field at all the excitation wavelength used with incident angles from 15° to 60° with a step size of 5° (*10*). The second term accounts for the Purcell effect, with $\lambda_s$ representing the scattered wavelength, Q being the quality factor (equal to 25 in our case), and V being the mode volume. The mode volume is calculated using the formula $V = \pi r^2 l$, where r is the beam radius and l is cavity path length which is 140 nm, which is the approximate upper limit to the mode volume with metal mirrors. The third term is the solid angle factor where the $\omega_0$ is the beam waist, is equal to 2r.

The enhancement factor for pentacene thin film Raman signal in the main text was calculated by comparing the amplitude of 1374 cm$^{-1}$ mode of a pentacene thin film (27 nm) on the mirror with a pentacene thin film (56 nm) on a glass slide using the formula given by equation (6) for 15°, 30°, and 45° measurement:

$$enhancement\ factor = \frac{(amplitude)_{mirror}}{(amplitude)_{glass}} * \frac{56}{27} \qquad (6)$$

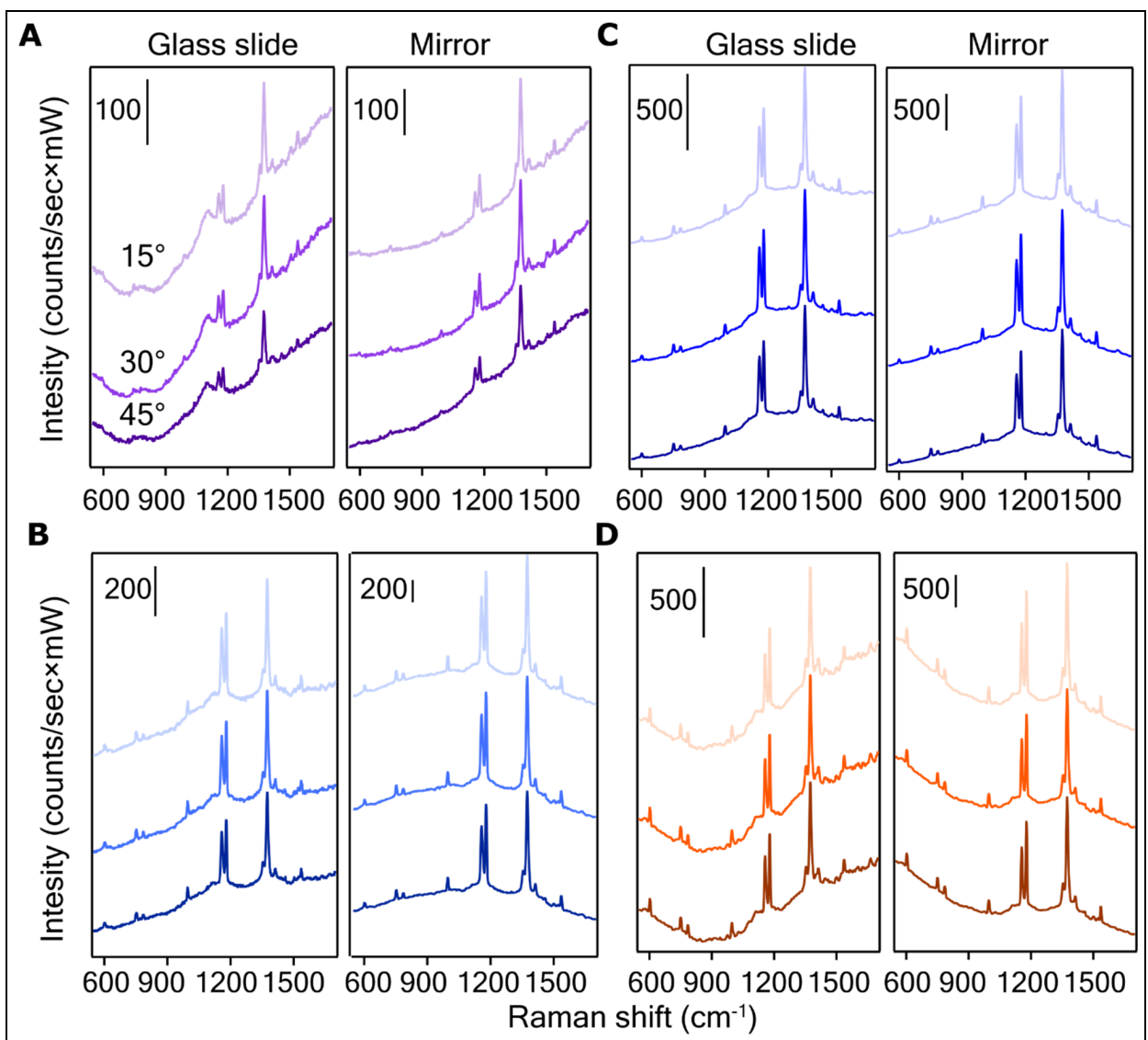

**Fig. S3.** Resonance Raman spectra of a pentacene thin film on a mirror (27nm) and on a glass slide (56 nm), measured at excitation energies of 2.09 eV **(A)**, 1.96 eV **(B)**, 1.94 eV **(C)**, and 1.88 eV **(D)**. In each subplot, the top, middle, and bottom spectra correspond to measurements at 15°, 30°, and 45°, respectively.

## 1.5 Pentacene thin film and cavity RRIA

The experimental Raman cross-sections ($\sigma$) were determined using equation (7) (*4*):

$$\sigma_{Raman} = \frac{8\pi}{3}\frac{(1+2\rho)}{(1+\rho)}\left[\left(\frac{\nu_L - \nu_{pc}}{\nu_L - \nu_{std}}\right)^3 \left(\frac{A_{pc}}{A_{std}}\right)\left(\frac{C_{std}}{C_{pc}}\right)\left(\frac{\partial \sigma_{std}}{\partial \Omega}\right)_{||\perp}\right] \quad (7)$$

where $\nu_{pc}$ and $\nu_{std}$ are the vibrational frequencies of pentacene and the 801 cm$^{-1}$ mode of cyclohexane, $A_{pc}$ and $A_{std}$ are the integrated areas of the vibrational mode of pentacene and the 801 cm$^{-1}$ cyclohexane mode, $C_{std}$ and $C_{pc}$ are the concentrations of pentacene and cyclohexane and $\left(\frac{\partial \sigma_{std}}{\partial \Omega}\right)_{||\perp}$ is the differential Raman cross-section calculated by extrapolating the value measured at 532 nm excitation (*5*). $C_{std}$ and $C_{pc}$ were calculated using the density of cyclohexane and pentacene. Resonance Raman intensities were corrected for self- absorption and enhancement factor (supplementary material 1.5) in both pentacene thin film and cavity.

The experimental absorption cross section ($\alpha$) was calculated using equation (8)

$$\alpha = \frac{2303 * Abs}{N_A * C_{pc} * L} \quad (8)$$

Abs is calculated using -log(reflectivity), $N_A$ is the Avagadro's number $C_{pc}$ is the pentacene concentration and L is the pentacene thickness. The absorption from metal mirrors is assumed to be negligible.

We simulate the absorption and Raman excitation profile using a simulator that was capable of handling up to two electronic states developed by Myers-Kelley et al. (*6*). To fit the experimental data with calculated data and extract the displacement ($\Delta$) value, various parameters such as peak position, homogeneous and inhomogeneous linewidths, experimental Raman cross-sections and Frank Condon (FC) displacement ($\Delta$) were iteratively adjusted and are shown in Table S2-S4. In these tables, E represents the energy, $\Gamma$ is the homogenous linewidth, M is the transition length and $\theta$ is the standard deviation of the inhomogeneous broadening. The subscripts 1 and 2 stands for the first and second electronic state for thin film and polaritonic state (LP and MP1) for the cavities.

Table S2. Parameters used to model absorbance and Raman excitation profile for pentacene thin film..

|  | Angle (°) | | |
| --- | --- | --- | --- |
|  | 15° | 30° | 45° |
| $E_1$ (cm$^{-1}$) | 15050 | 15050 | 15050 |
| $\Gamma_1$ (cm$^{-1}$) | 25 | 25 | 16 |
| $M_1$ (Å) | 1.07 | 1.08 | 1.07 |
| $E_2$ (cm$^{-1}$) | 17100 | 17100 | 17100 |
| $\Gamma_2$ (cm$^{-1}$) | 700 | 700 | 700 |
| $M_2$ (Å) | 0.55 | 0.55 | 0.52 |
| $\theta$ (cm$^{-1}$) | 310 | 310 | 310 |

**Table S3.** Parameters used to model absorbance and Raman excitation profiles for 27 nm pentacene cavity.

| | Angle (°) | | | | | | | | | |
|---|---|---|---|---|---|---|---|---|---|---|
| | 15° | 20° | 25° | 30° | 35° | 40° | 45° | 50° | 55° | 60° |
| $E_1$ (cm$^{-1}$) | 14370 | 14410 | 14465 | 14530 | 14600 | 14640 | 14670 | 14685 | 14700 | 14700 |
| $\Gamma_1$ (cm$^{-1}$) | 300 | 350 | 250 | 180 | 110 | 80 | 70 | 30 | 20 | 15 |
| $M_1$ (Å) | 1.01 | 0.96 | 0.92 | 0.94 | 0.86 | 0.77 | 0.71 | 0.54 | 0.47 | 0.41 |
| $E_2$ (cm$^{-1}$) | 15306 | 15330 | 15400 | 15500 | 15630 | 15710 | 15800 | 15850 | 15900 | 15900 |
| $\Gamma_2$ (cm$^{-1}$) | 500 | 530 | 450 | 430 | 400 | 600 | 650 | 700 | 800 | 900 |
| $M_2$ (Å) | 0.81 | 0.81 | 0.82 | 0.85 | 0.82 | 0.83 | 0.74 | 0.66 | 0.59 | 0.53 |
| $\theta$ (cm$^{-1}$) | 90 | 48 | 120 | 200 | 230 | 250 | 250 | 220 | 220 | 220 |

Table S4. Parameters used to model absorbance and Raman excitation profile for 39 nm pentacene cavity.

| | Angle (°) | | | | | | | | | |
|---|---|---|---|---|---|---|---|---|---|---|
| | 15° | 20° | 25° | 30° | 35° | 40° | 45° | 50° | 55° | 60° |
| $E_1$ (cm$^{-1}$) | 14300 | 14350 | 14390 | 14440 | 14490 | 14540 | 14600 | 14630 | 14660 | 14690 |
| $\Gamma_1$ (cm$^{-1}$) | 250 | 250 | 200 | 150 | 110 | 90 | 60 | 40 | 24 | 13 |
| $M_1$ (Å) | 0.78 | 0.74 | 0.73 | 0.68 | 0.63 | 0.61 | 0.53 | 0.47 | 0.40 | 0.33 |
| $E_2$ (cm$^{-1}$) | 15400 | 15400 | 15450 | 15500 | 15610 | 15630 | 15700 | 15740 | 15750 | 15780 |
| $\Gamma_2$ (cm$^{-1}$) | 450 | 500 | 500 | 550 | 600 | 600 | 650 | 650 | 700 | 750 |
| $M_2$ (Å) | 0.57 | 0.59 | 0.60 | 0.58 | 0.57 | 0.52 | 0.48 | 0.43 | 0.36 | 0.30 |
| $\theta$ (cm$^{-1}$) | 100 | 100 | 150 | 170 | 200 | 250 | 260 | 270 | 240 | 230 |

### 1.6 **Computational Methods**

The ground state geometry was optimized and then frequency calculations were performed using density functional theory (DFT) in ORCA and a def2-SVP basis with B3LYP functional was used for these calculations. The Franck-Condon relaxed excited state structure was determined using a freely available Python script (*11*). This program calculates the excited state geometry by projecting dimensionless normal mode displacements onto cartesian coordinates. To utilize script, the following input files were required: normal mode displacement vectors, normal mode frequencies, dimensionless normal mode displacements, Cartesian coordinate structure of the molecule, and atomic masses which can be extracted from the ORCA output file. By specifying four atoms defining the internal coordinate of interest, the program calculates the total displacement along all possible internal coordinates (dihedrals, angles, and bonds). Additionally, it generates individual files containing the internal coordinate displacement projected onto normal coordinates.

### 2. **MP1 Δ values**

There are three peaks in the REP: the first and second are influenced mostly by changes in the LP Δ values, while the third peak is affected by changes in the MP1 Δ values. For low angles and low-frequency modes, we do have a certain Δ value for both LP and MP. However, at high angles and high frequencies, we lack data points to fit the MP1 Δ value. Therefore, we cannot confidently comment on the Δ values of MP1 state and need more data points for proper interpretation. Figure S4 shows the MP1 Δ values for the 600 cm$^{-1}$, 754 cm$^{-1}$, and 784 cm$^{-1}$ mode. For angles greater than or equal to 35°, the 600 cm$^{-1}$ mode shows a significant variability in its Δ value. In contrast, the 754 cm$^{-1}$ and 784 cm$^{-1}$ modes, appear to maintain a $\Delta_{MP1}=0$ value in this angle range. Rest of the modes seems to have a $\Delta_{MP1}=0$ value across all angles examined in the angle-resolved RRIA analysis. However, due to the limited number of data points for these higher frequency modes, a precise determination of their MP1 Δ values is not possible in our analysis.

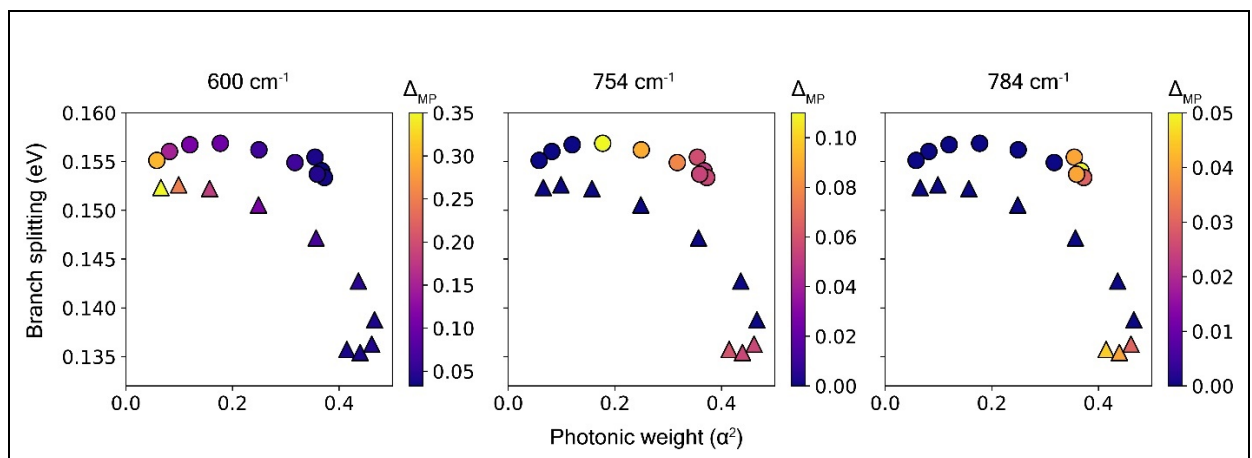

**Fig. S4.** Displacement ($\Delta_{MP}$) for the middle polaritonic state 1 (MP1) for 27 nm (triangle) and 39 nm (circle) pentacene cavities plotted against the branch splitting and photonic weight. The color of the point indicates the displacement value.

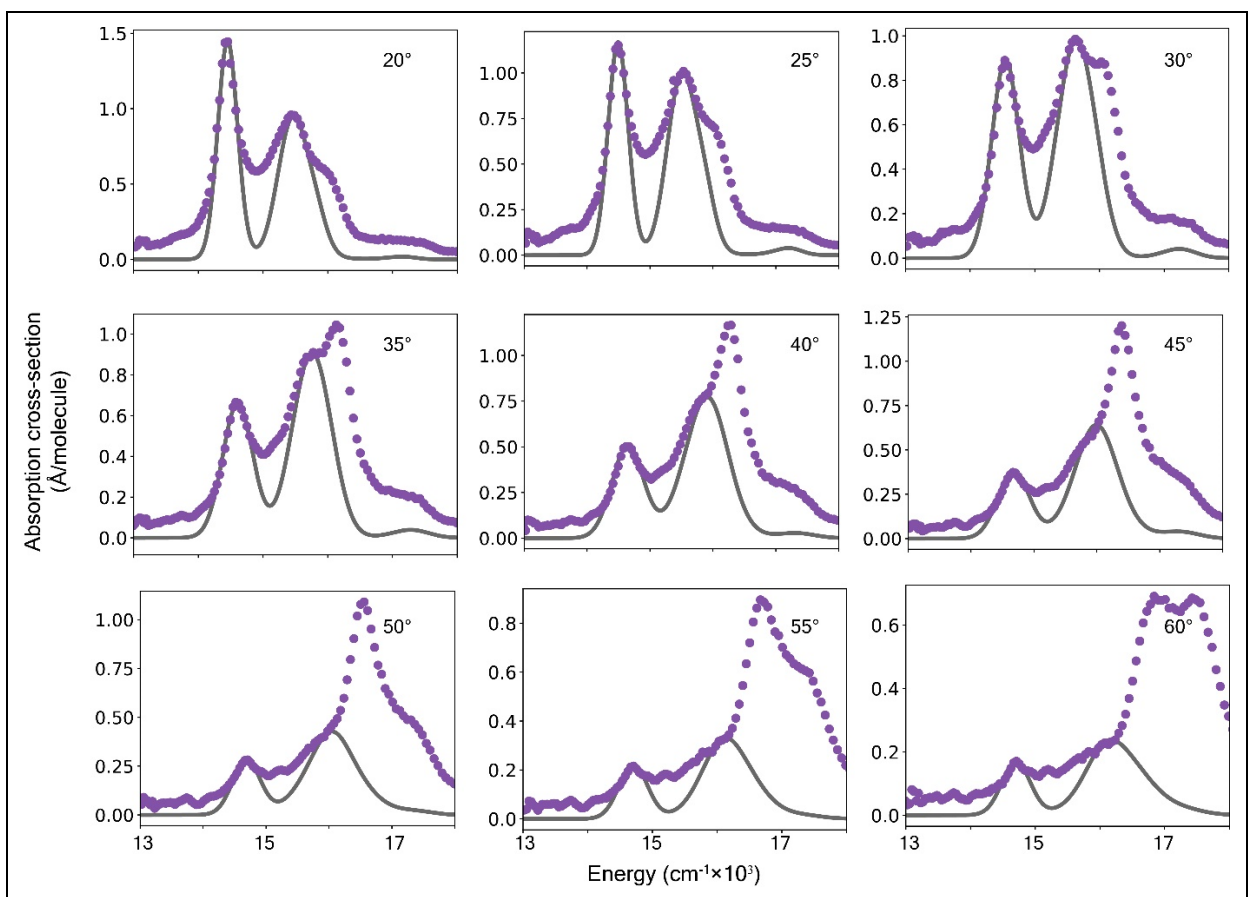

**Fig. S5.** Modeled (black lines) and experimental (purple points) absorption cross-sections of the 27 nm pentacene cavity at various angles.

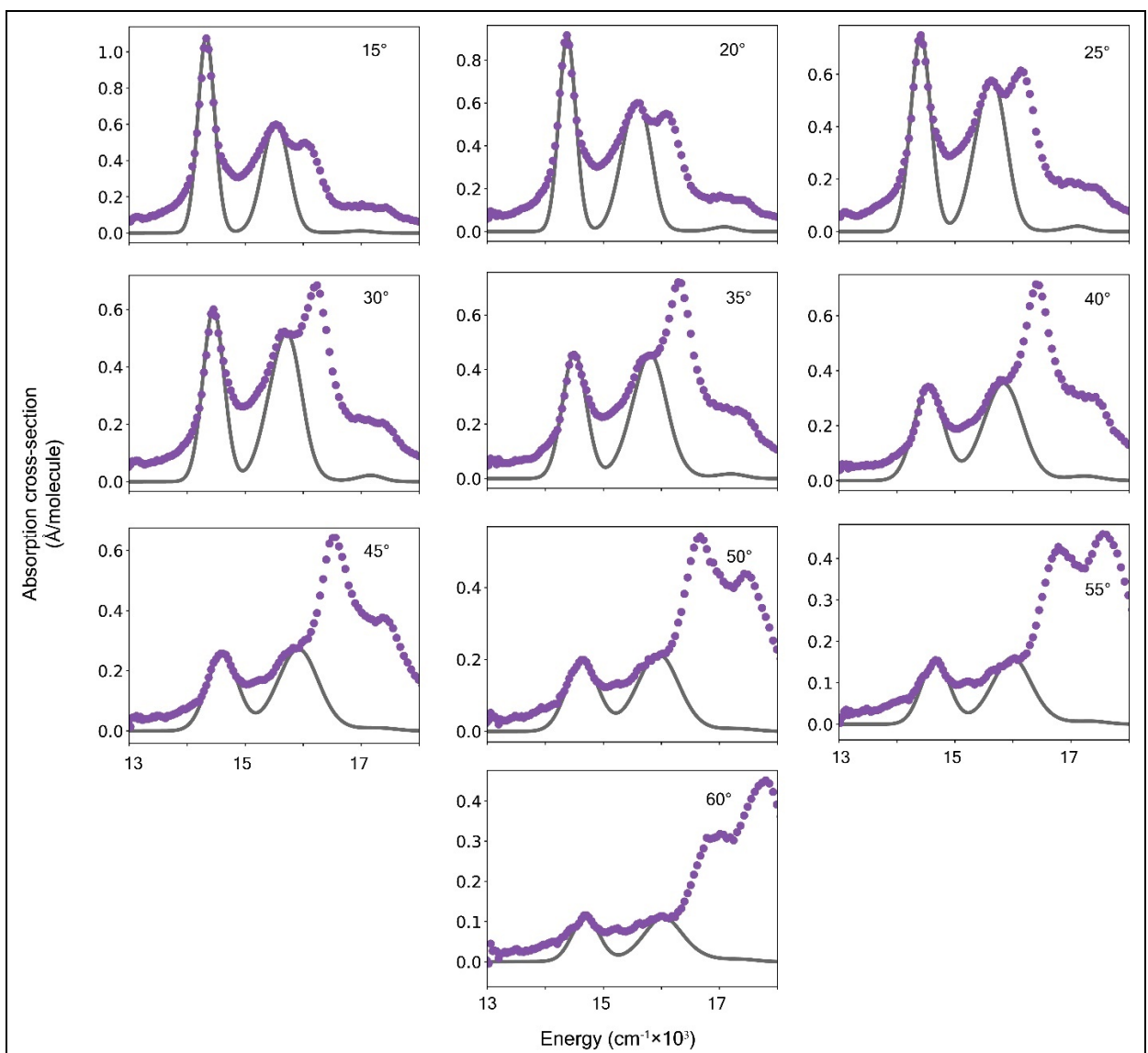

**Fig. S6.** Modeled (black lines) and experimental (purple points) absorption cross-sections of the 39 nm pentacene cavity at various angles.

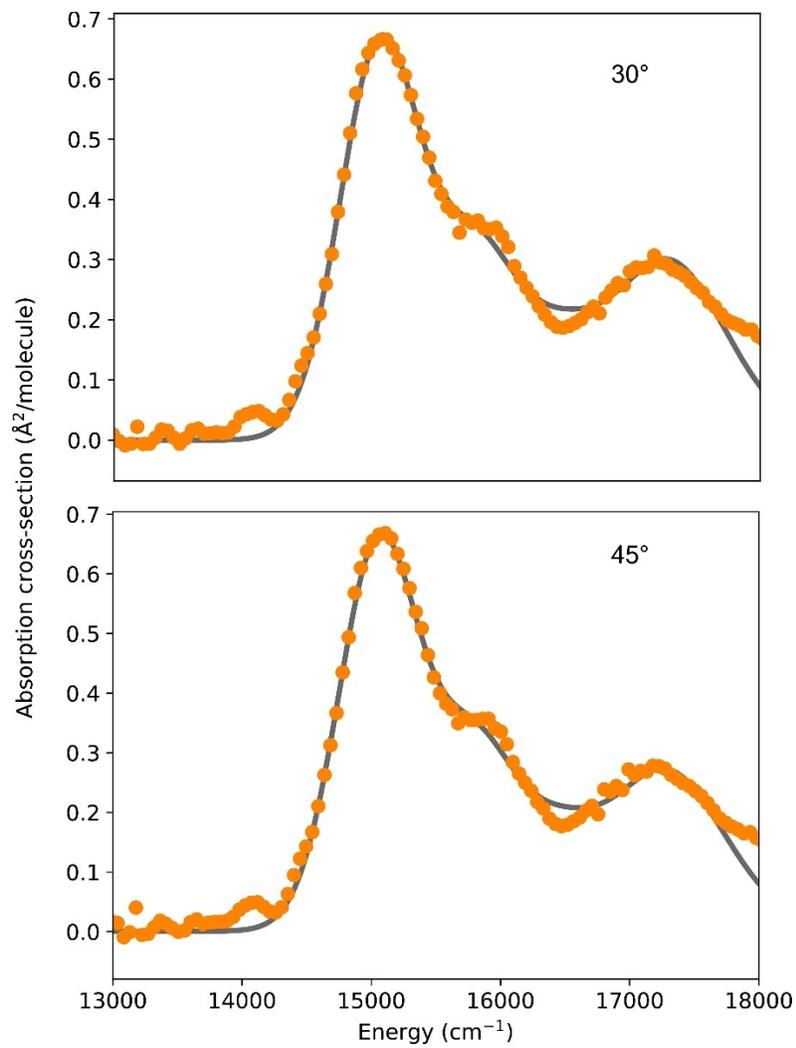

**Fig. S7.** Modeled (black lines) and experimental (orange points) absorption cross-sections of the 27 nm pentacene thin film at 30° and 45°.

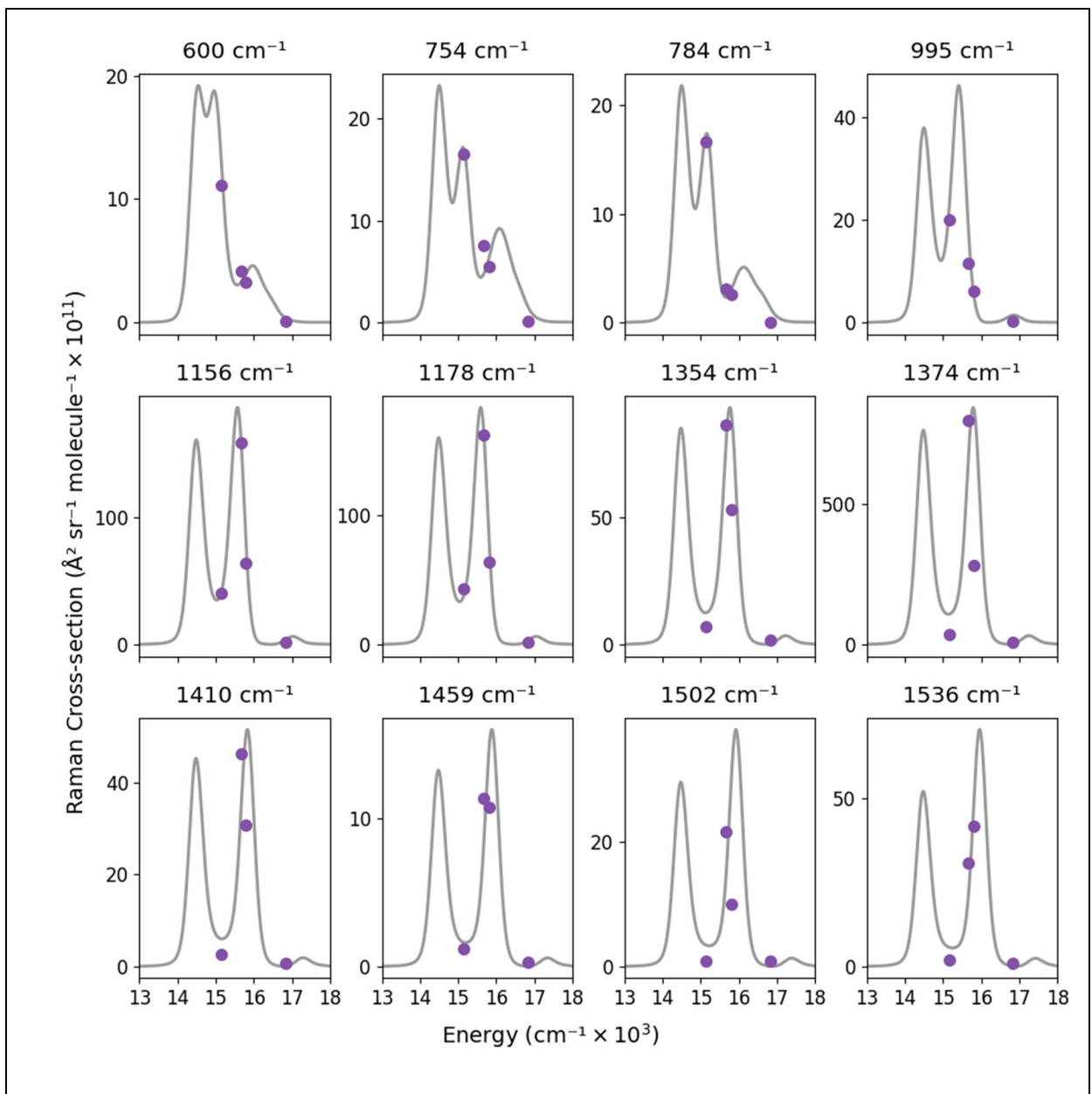

**Fig. S8.** Modeled (grey lines) and experimental (purple points) Raman excitation profiles for all Raman active vibrational modes of the 27 nm pentacene cavity at 20°.

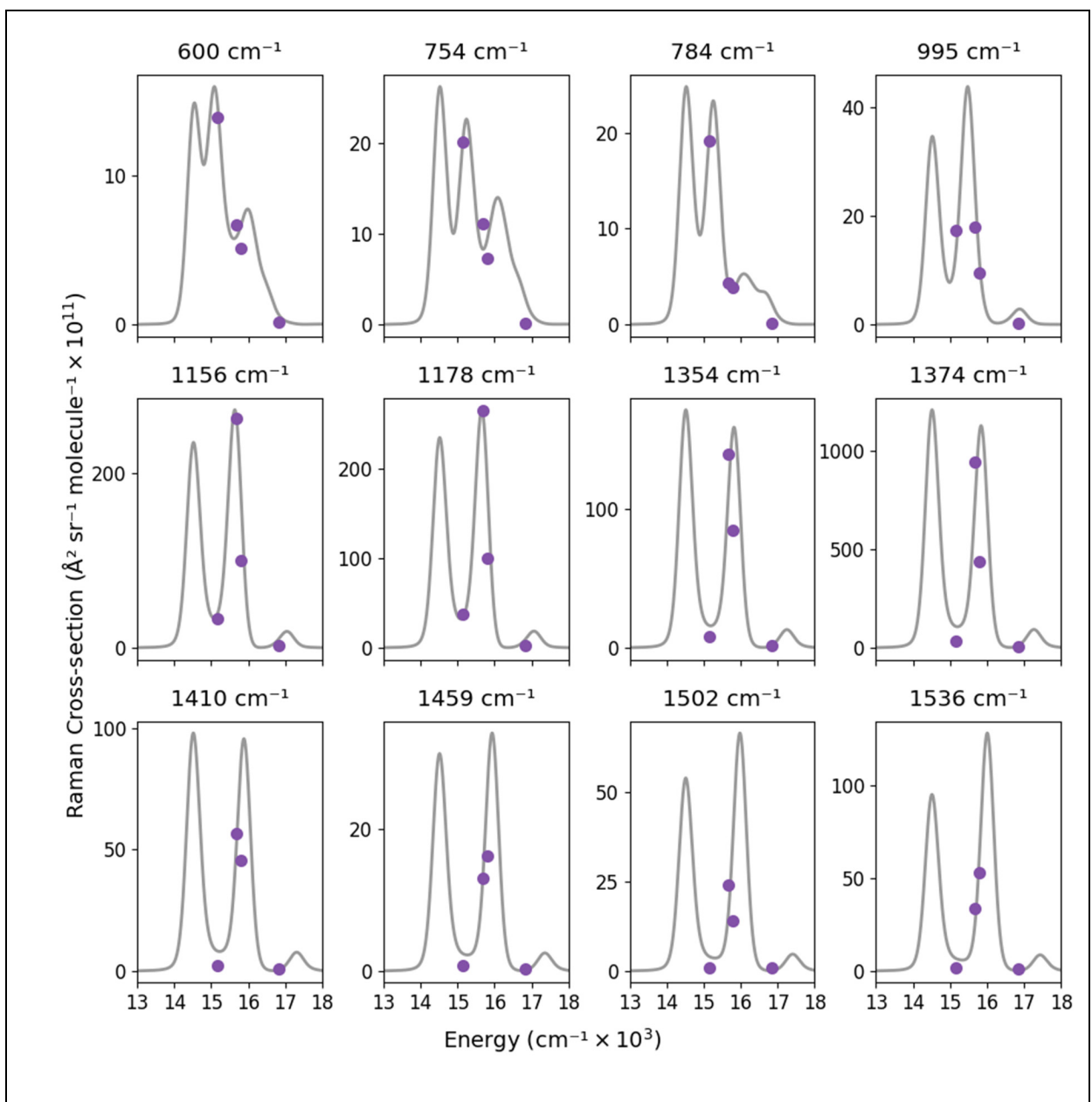

**Fig. S9.** Modeled (grey lines) and experimental (purple points) Raman excitation profiles for all Raman active vibrational modes of the 27 nm pentacene cavity at 25°.

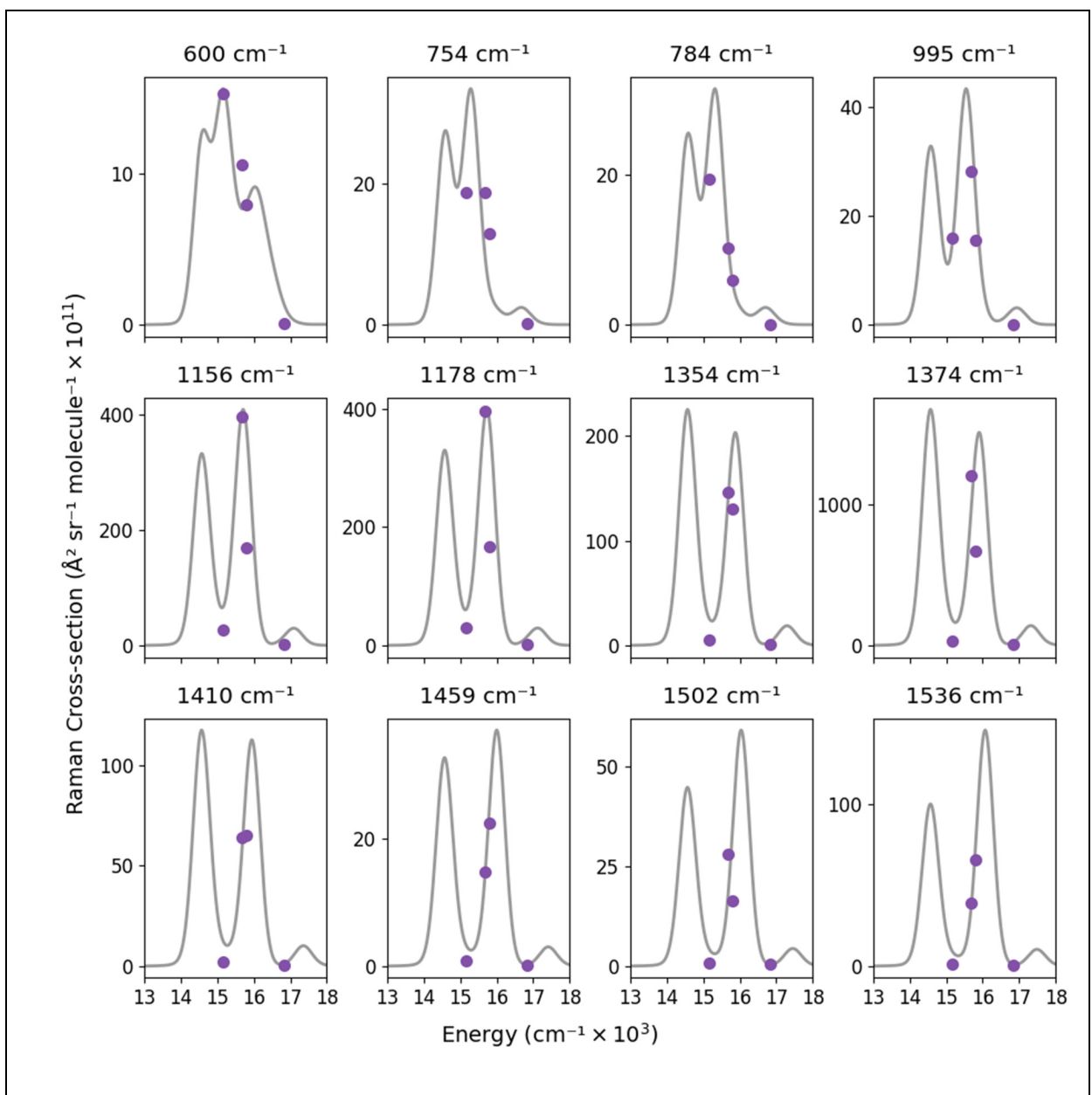

**Fig. S10.** Modeled (grey lines) and experimental (purple points) Raman excitation profiles for all Raman active vibrational modes of the 27 nm pentacene cavity at 30°.

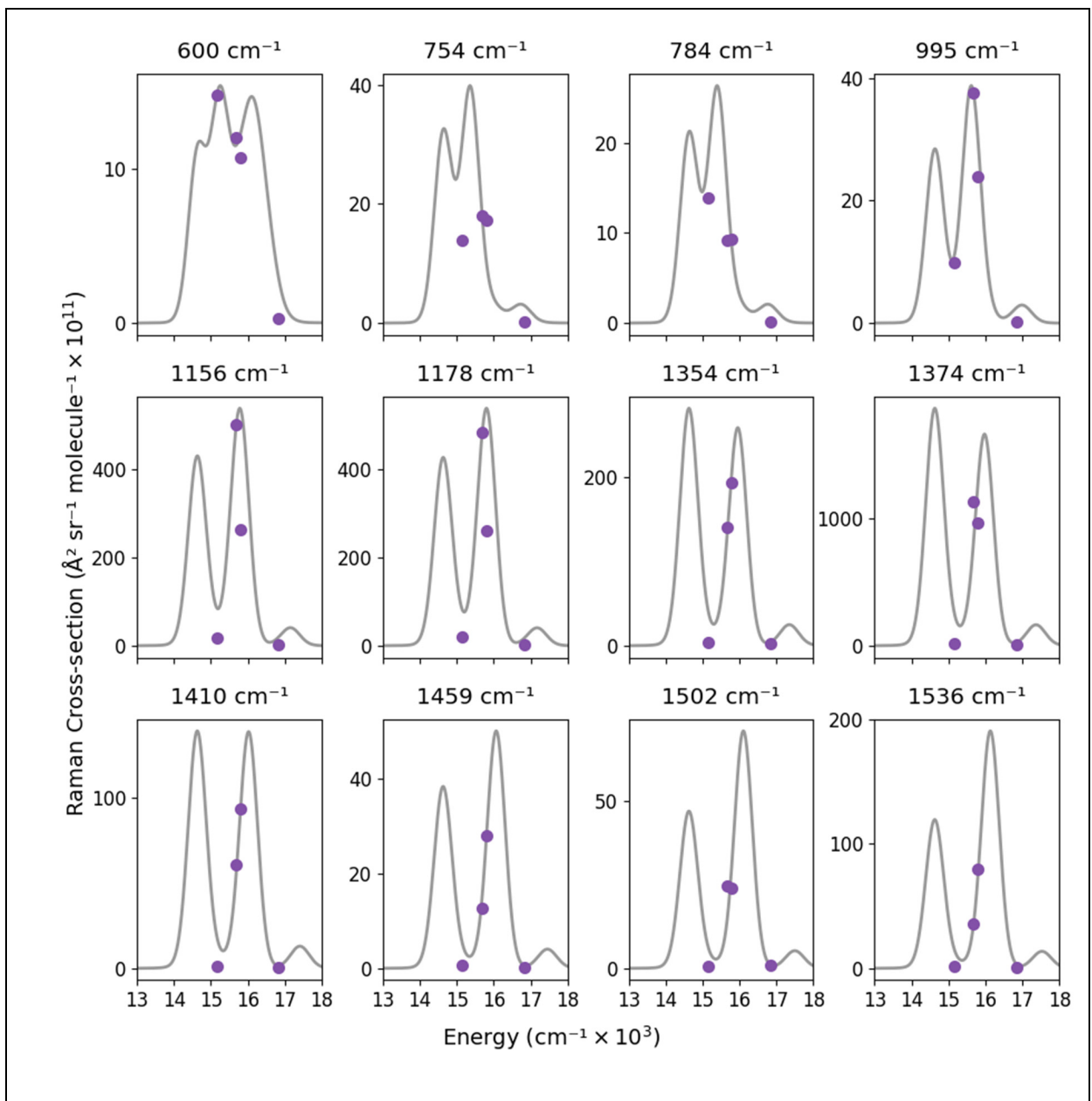

**Fig. S11.** Modeled (grey lines) and experimental (purple points) Raman excitation profiles for all Raman active vibrational modes of the 27 nm pentacene cavity at 35°.

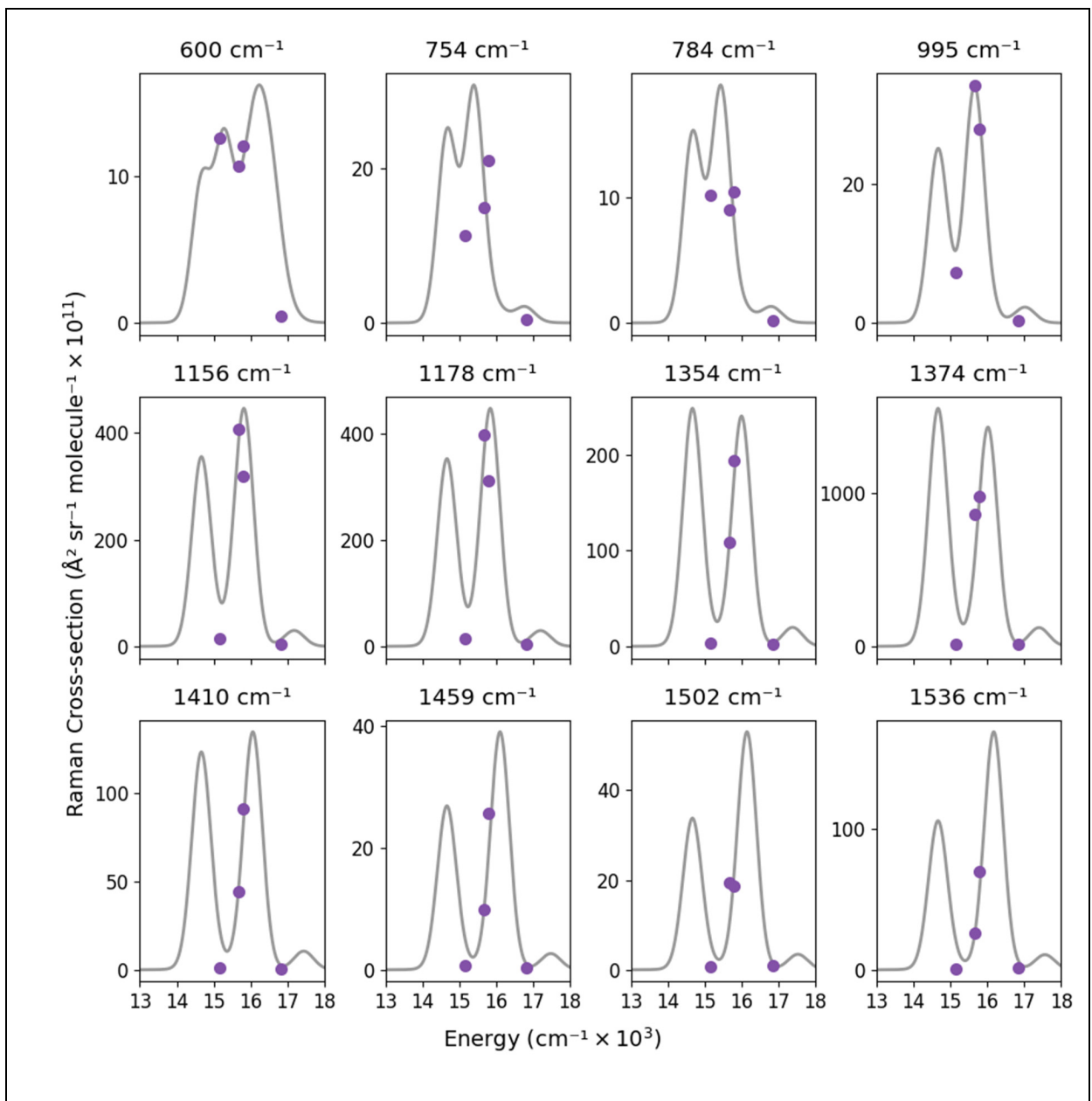

**Fig. S12.** Modeled (grey lines) and experimental (purple points) Raman excitation profiles for all Raman active vibrational modes of the 27 nm pentacene cavity at 40°.

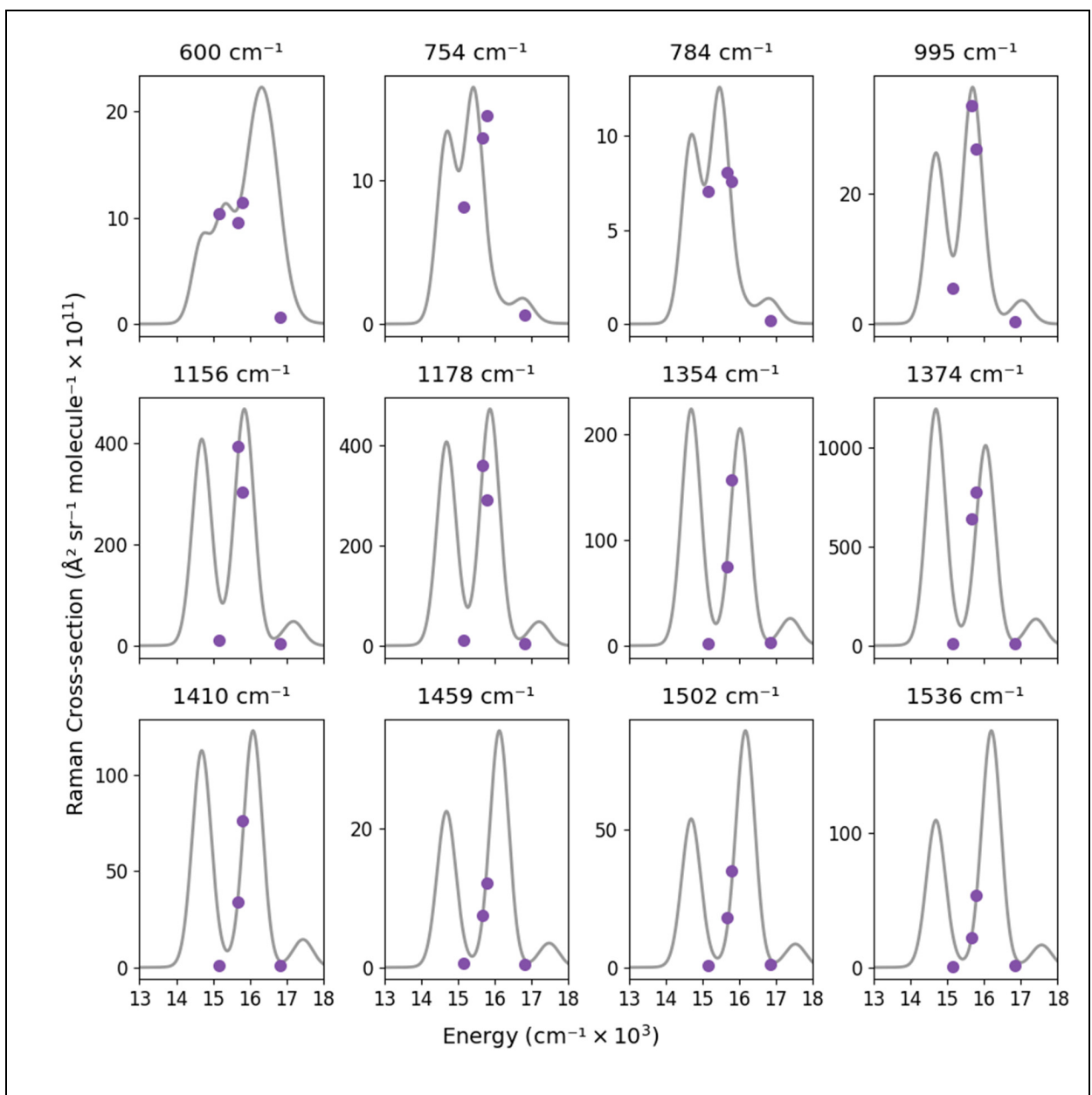

**Fig. S13.** Modeled (grey lines) and experimental (purple points) Raman excitation profiles for all Raman active vibrational modes of the 27 nm pentacene cavity at 45°.

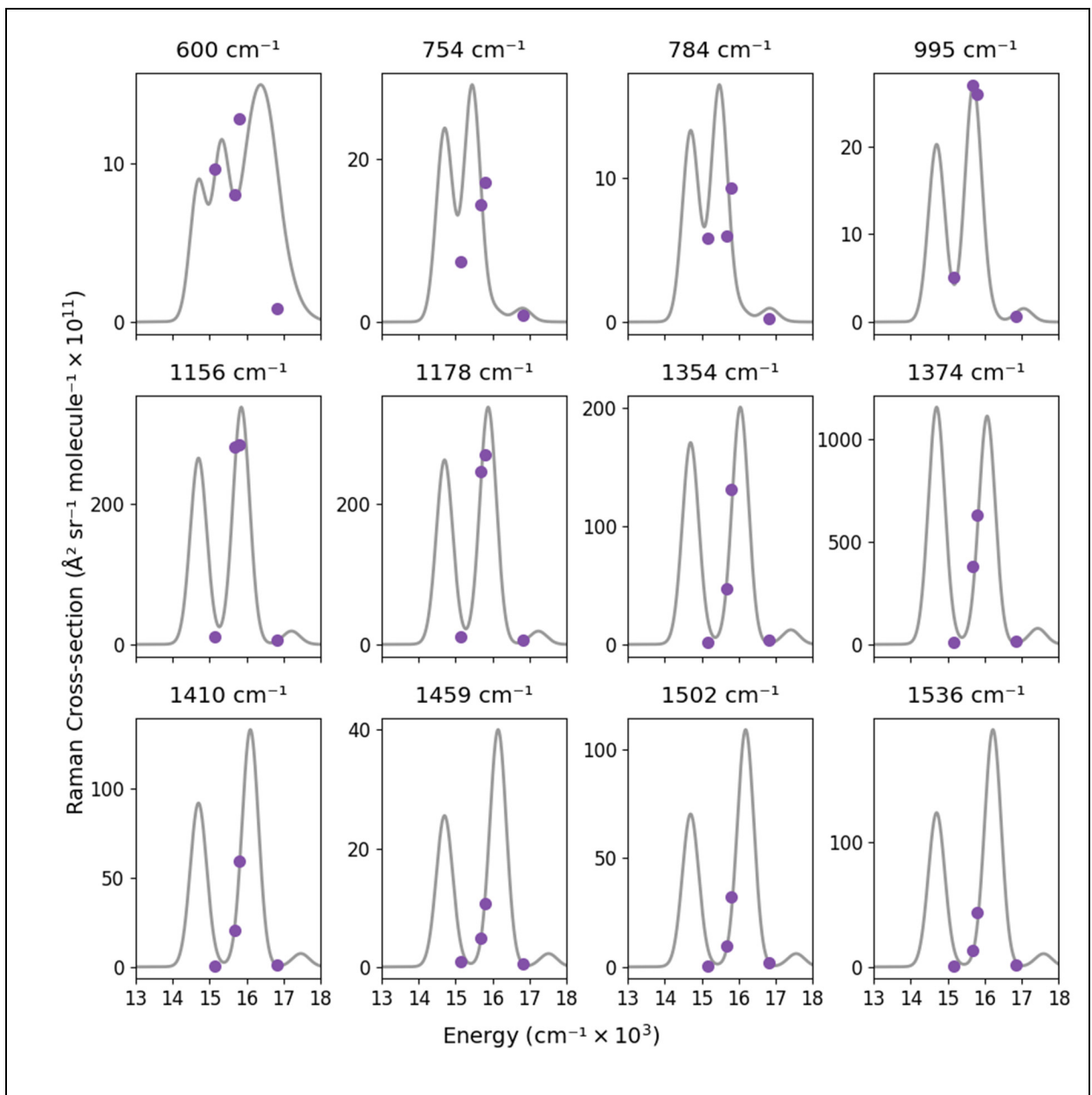

**Fig. S14.** Modeled (grey lines) and experimental (purple points) Raman excitation profiles for all Raman active vibrational modes of the 27 nm pentacene cavity at 50°.

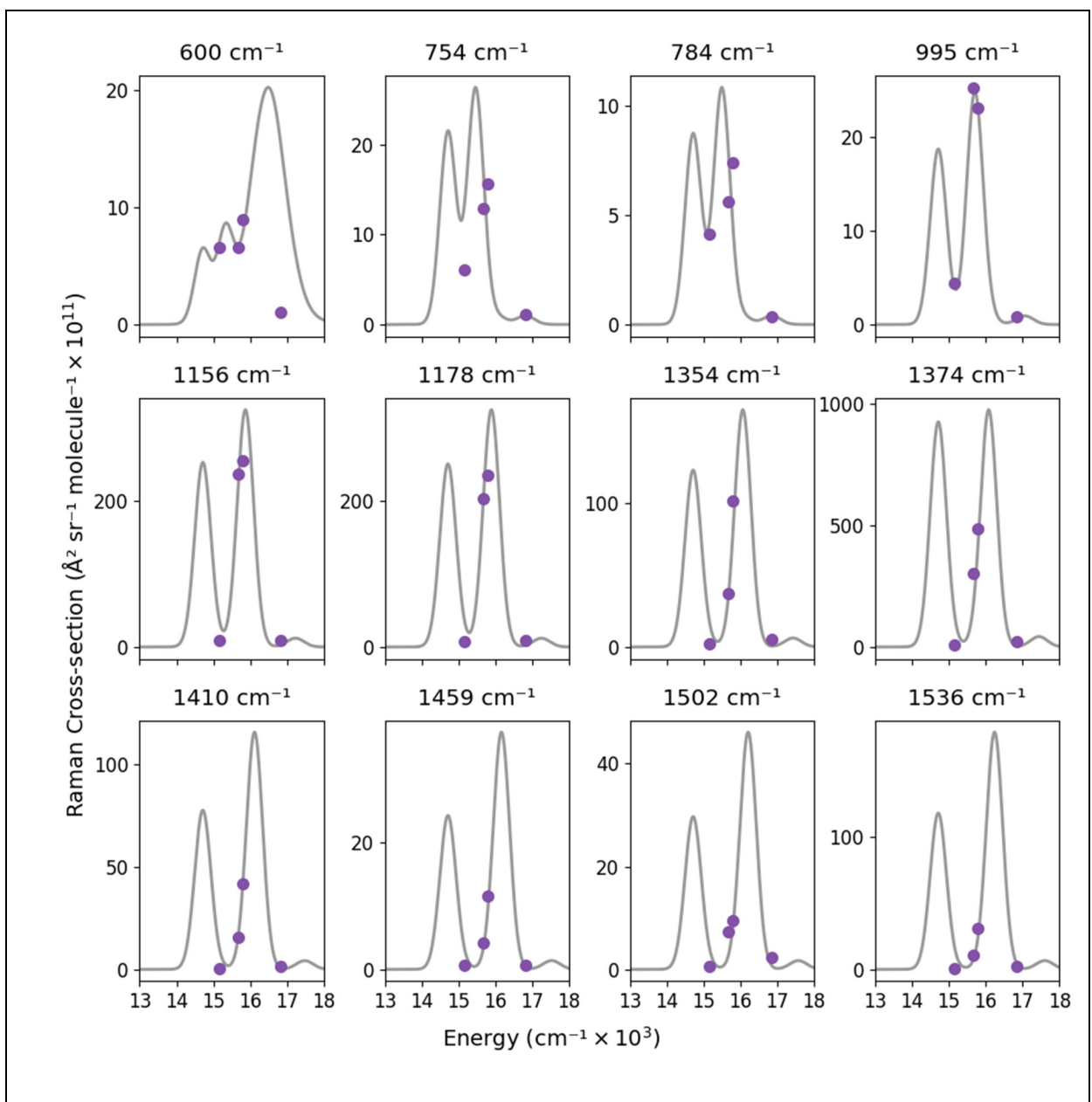

**Fig. S15.** Modeled (grey lines) and experimental (purple points) Raman excitation profiles for all Raman active vibrational modes of the 27 nm pentacene cavity at 55°.

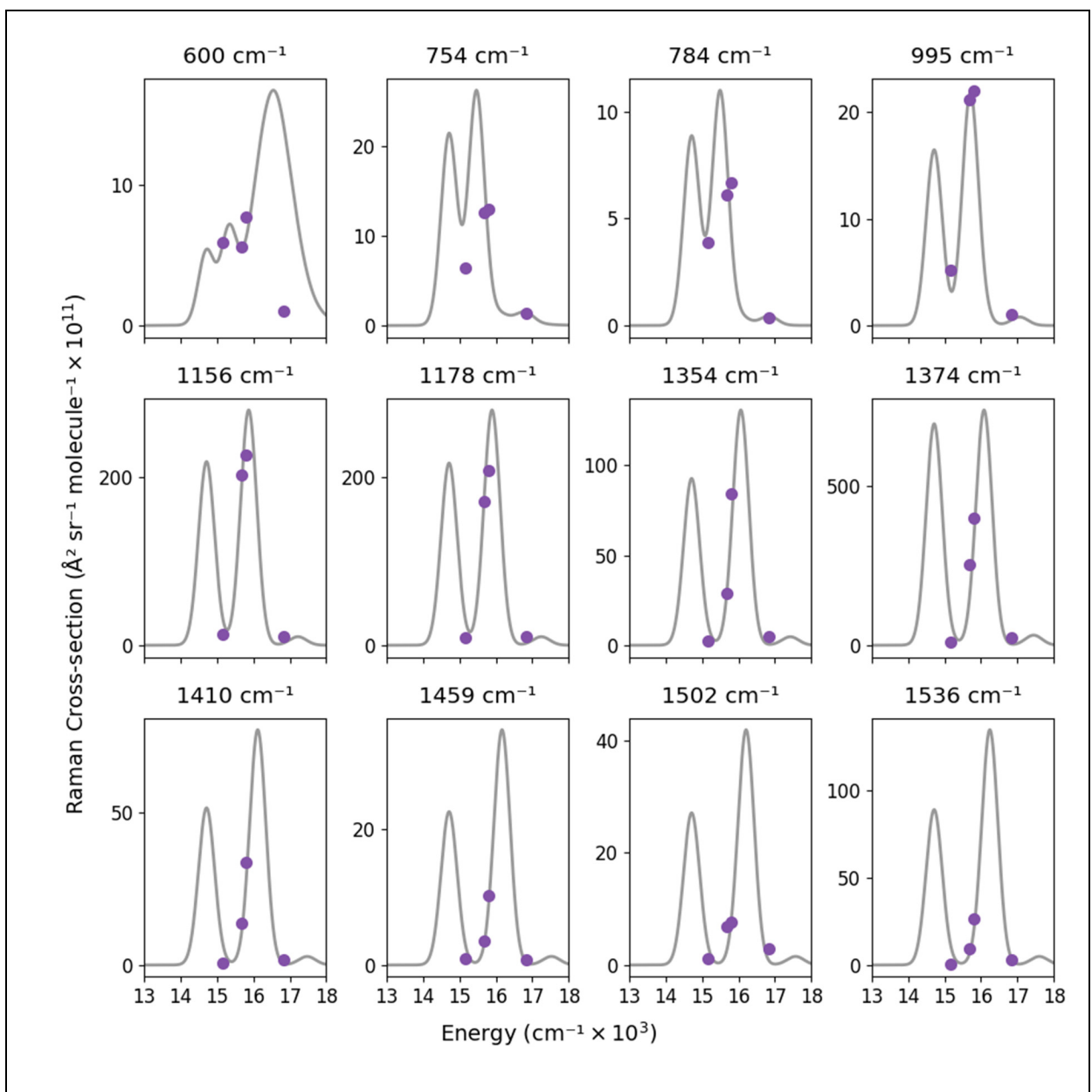

**Fig. S16.** Modeled (grey lines) and experimental (purple points) Raman excitation profiles for all Raman active vibrational modes of the 27 nm pentacene cavity at 60°.

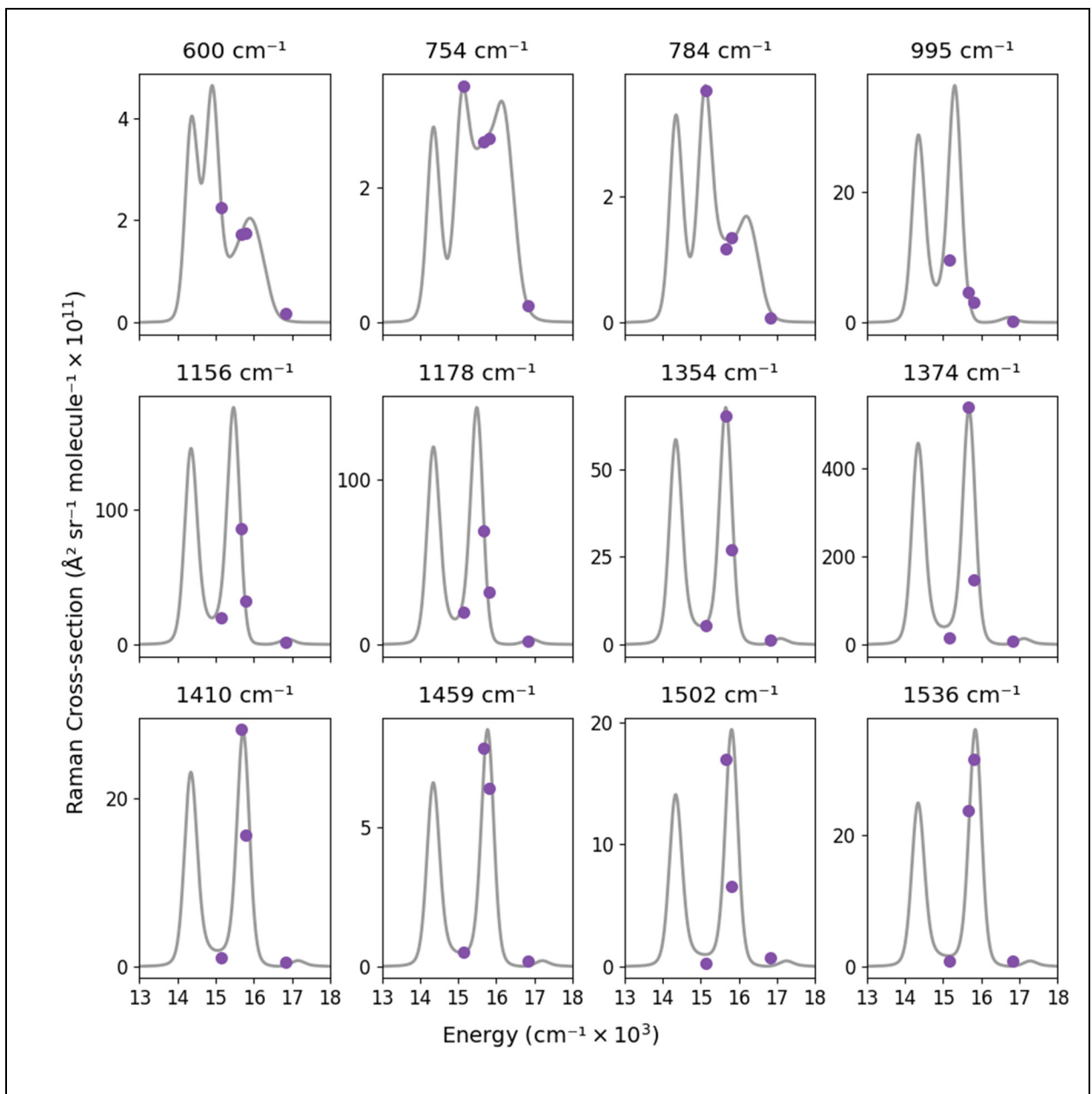

**Fig. S17.** Modeled (grey lines) and experimental (purple points) Raman excitation profiles for all Raman active vibrational modes of the 39 nm pentacene cavity at 15°.

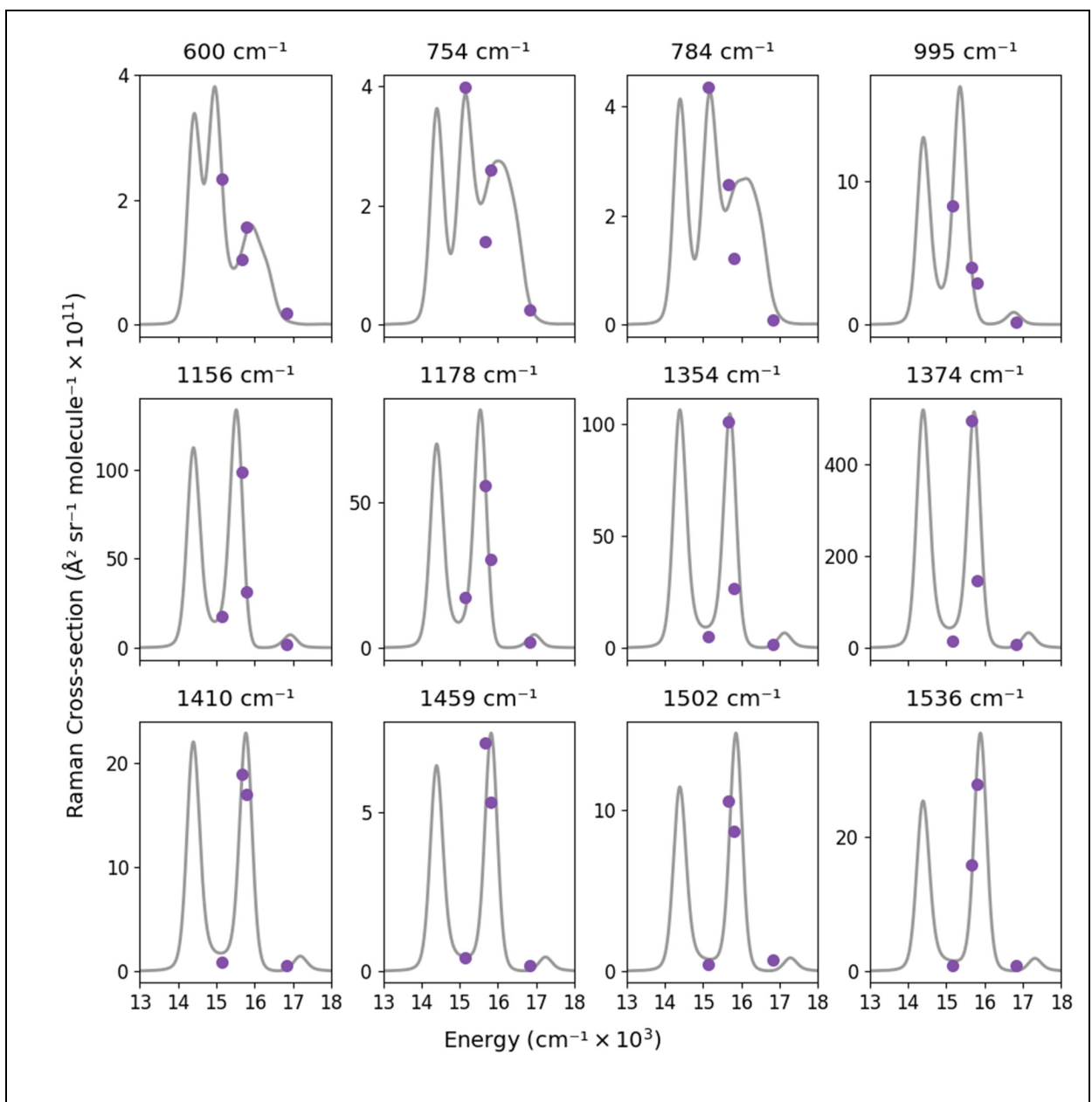

**Fig. S18.** Modeled (grey lines) and experimental (purple points) Raman excitation profiles for all Raman active vibrational modes of the 39 nm pentacene cavity at 20°.

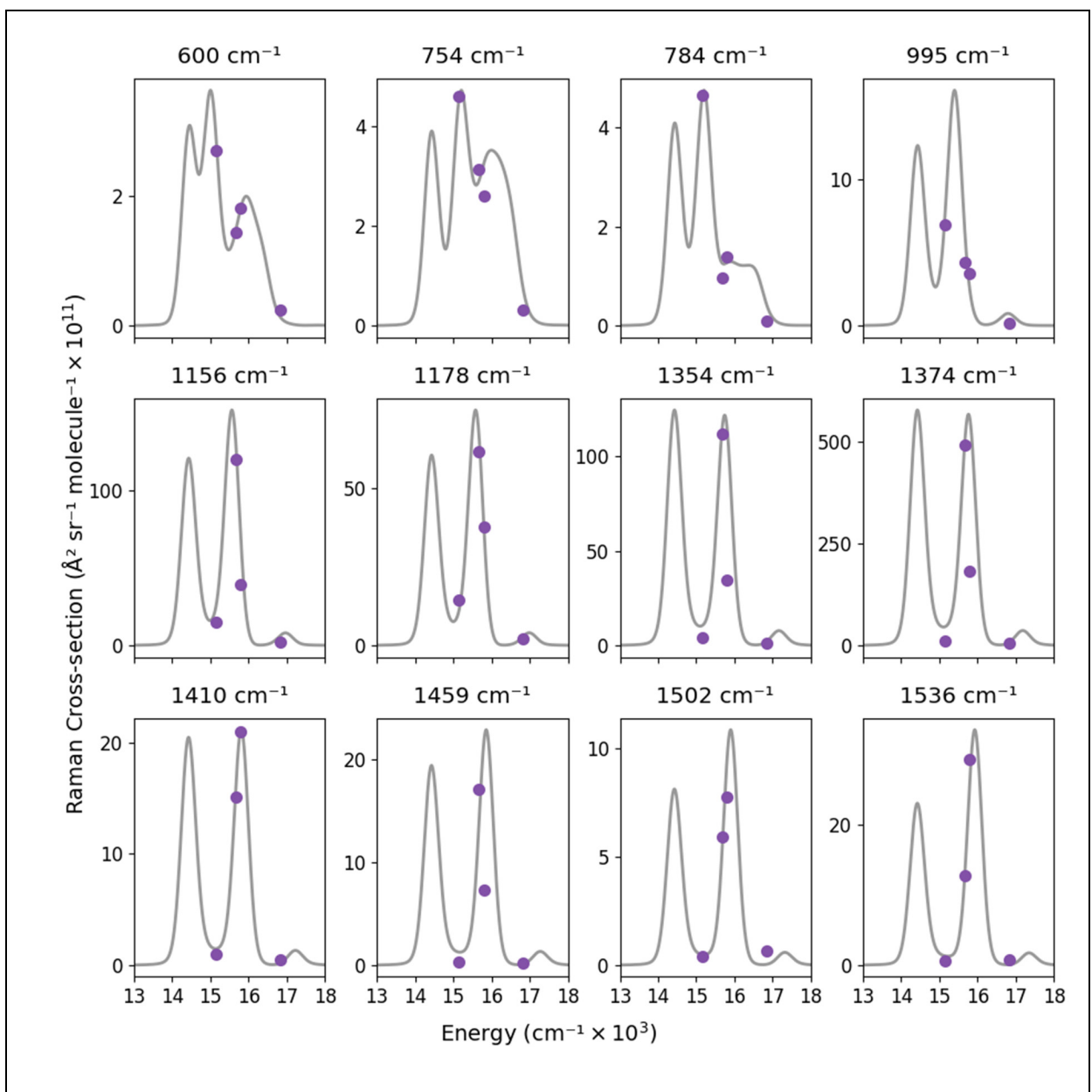

**Fig. S19.** Modeled (grey lines) and experimental (purple points) Raman excitation profiles for all Raman active vibrational modes of the 39 nm pentacene cavity at 25°.

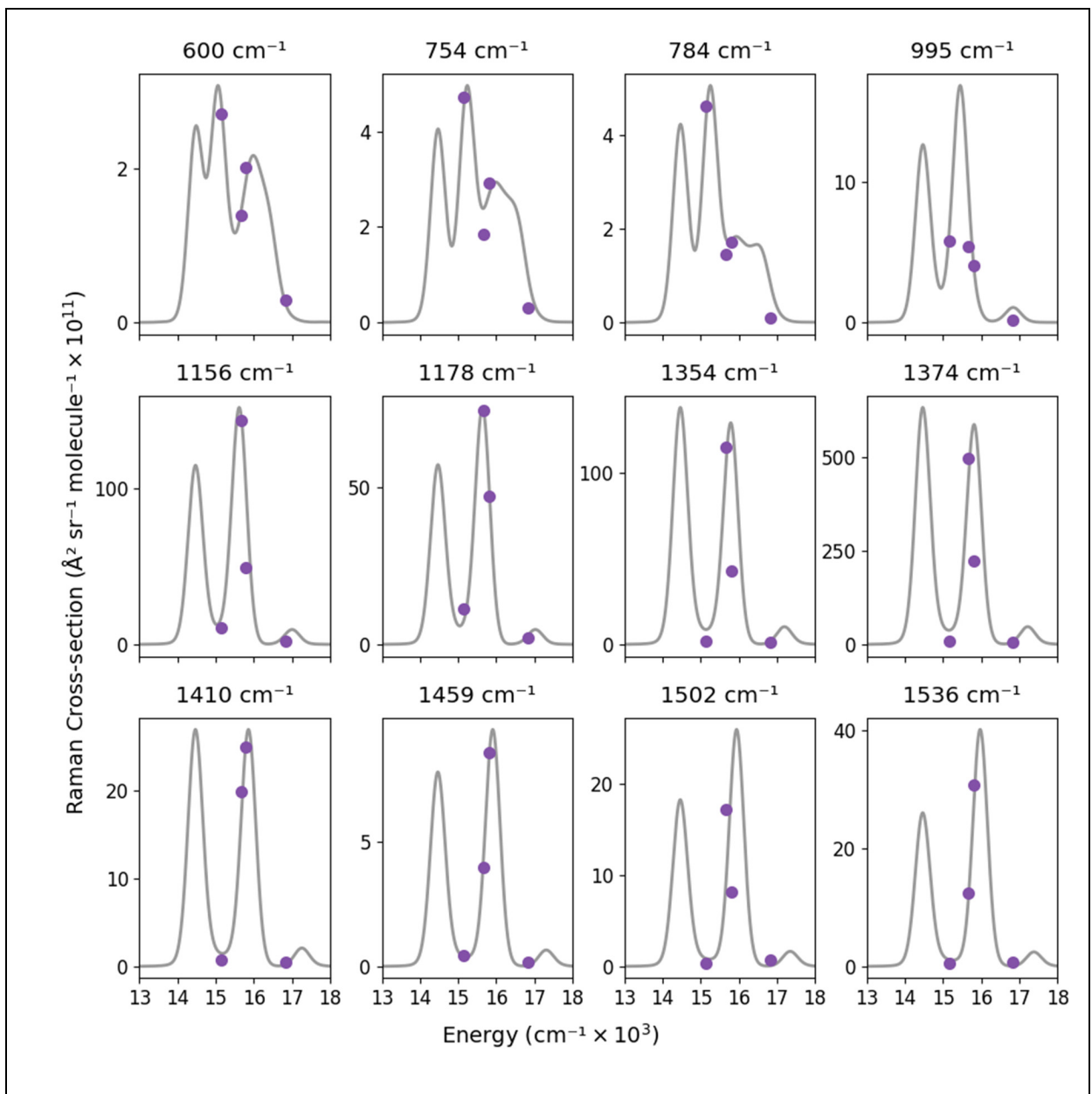

**Fig. S20.** Modeled (grey lines) and experimental (purple points) Raman excitation profiles for all Raman active vibrational modes of the 39 nm pentacene cavity at 30°.

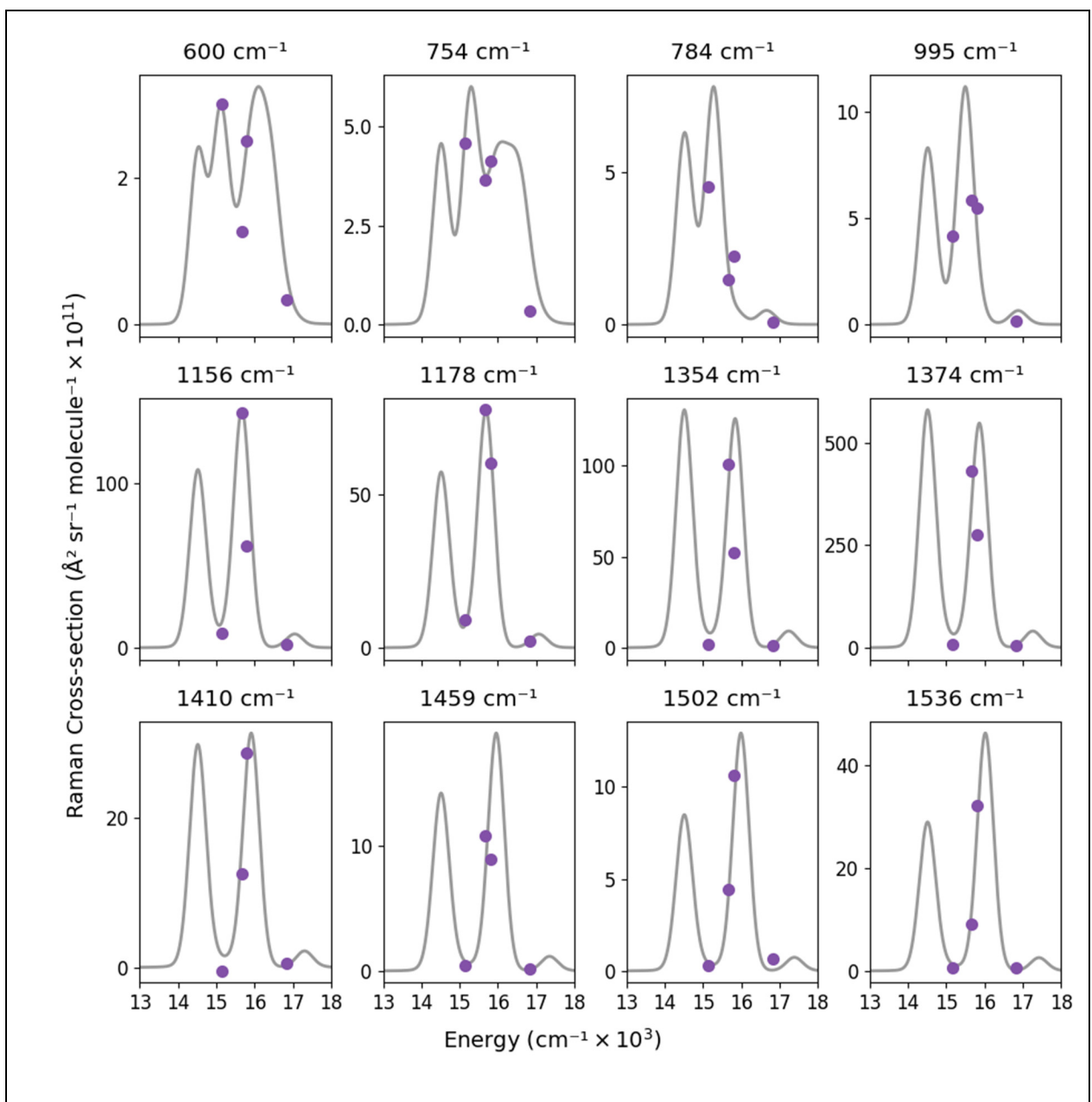

**Fig. S21.** Modeled (grey lines) and experimental (purple points) Raman excitation profiles for all Raman active vibrational modes of the 39 nm pentacene cavity at 35°.

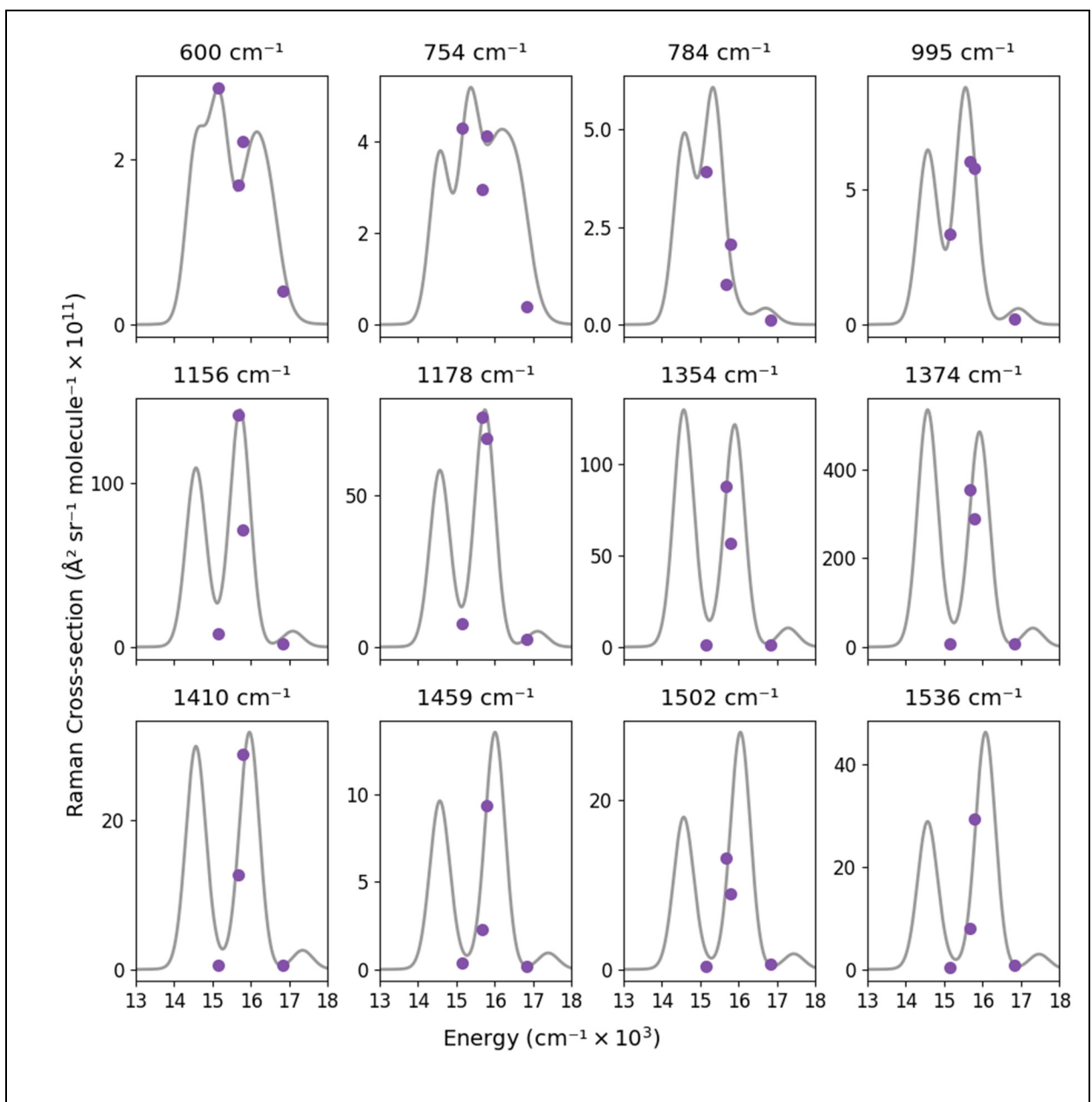

**Fig. S22.** Modeled (grey lines) and experimental (purple points) Raman excitation profiles for all Raman active vibrational modes of the 39 nm pentacene cavity at 40°.

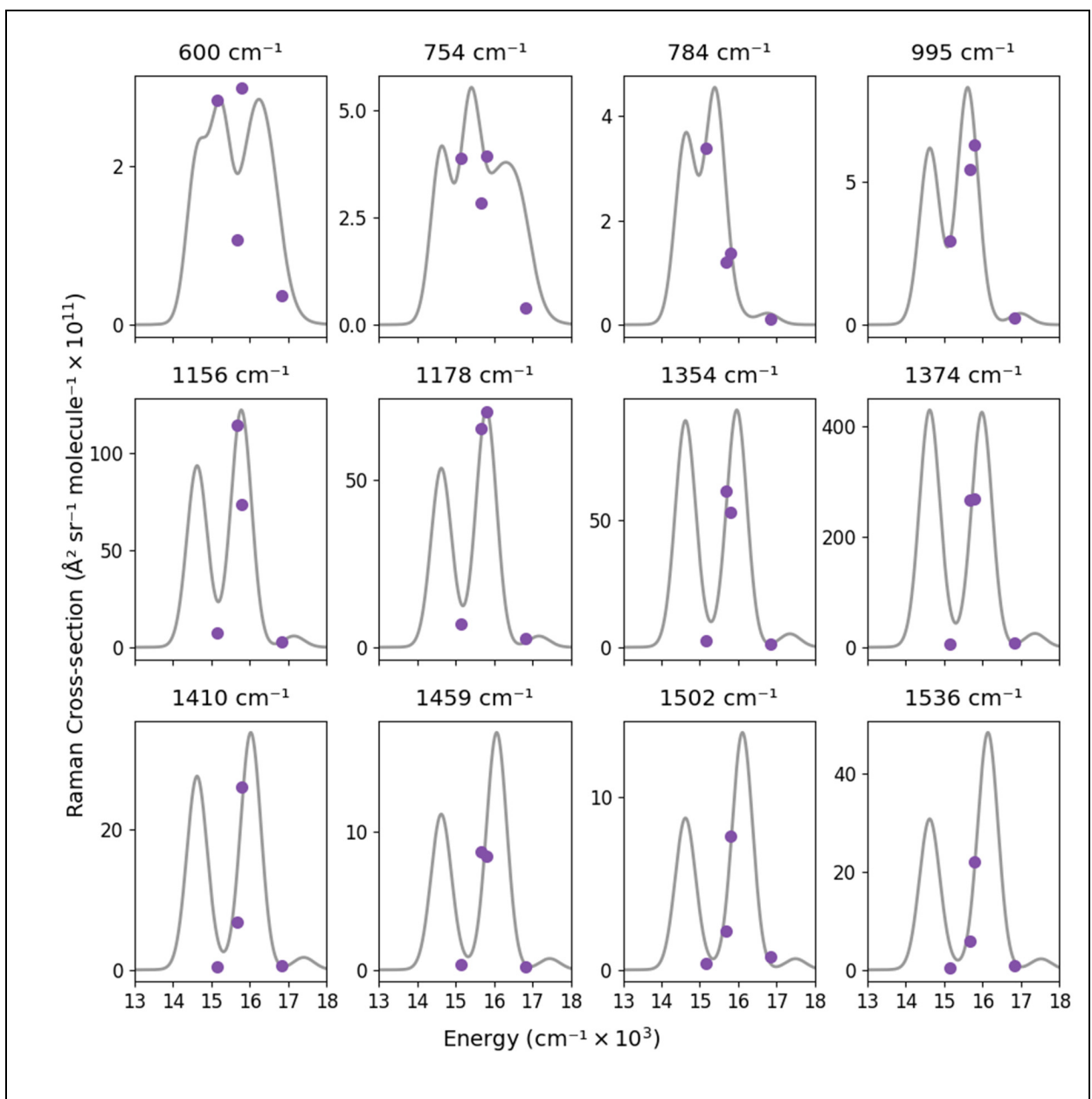

**Fig. S23.** Modeled (grey lines) and experimental (purple points) Raman excitation profiles for all Raman active vibrational modes of the 39 nm pentacene cavity at 45°.

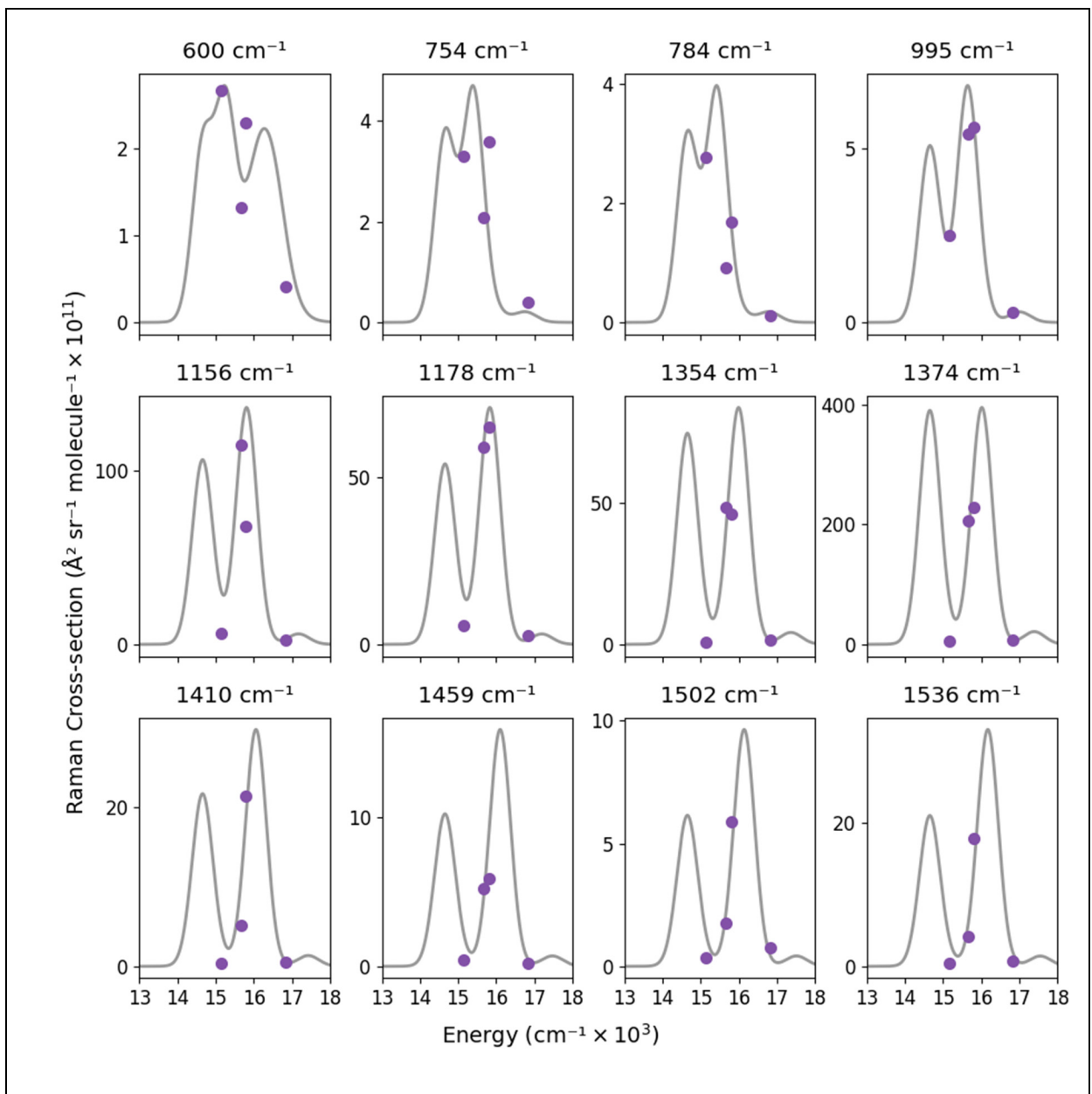

**Fig. S24.** Modeled (grey lines) and experimental (purple points) Raman excitation profiles for all Raman active vibrational modes of the 39 nm pentacene cavity at 50°.

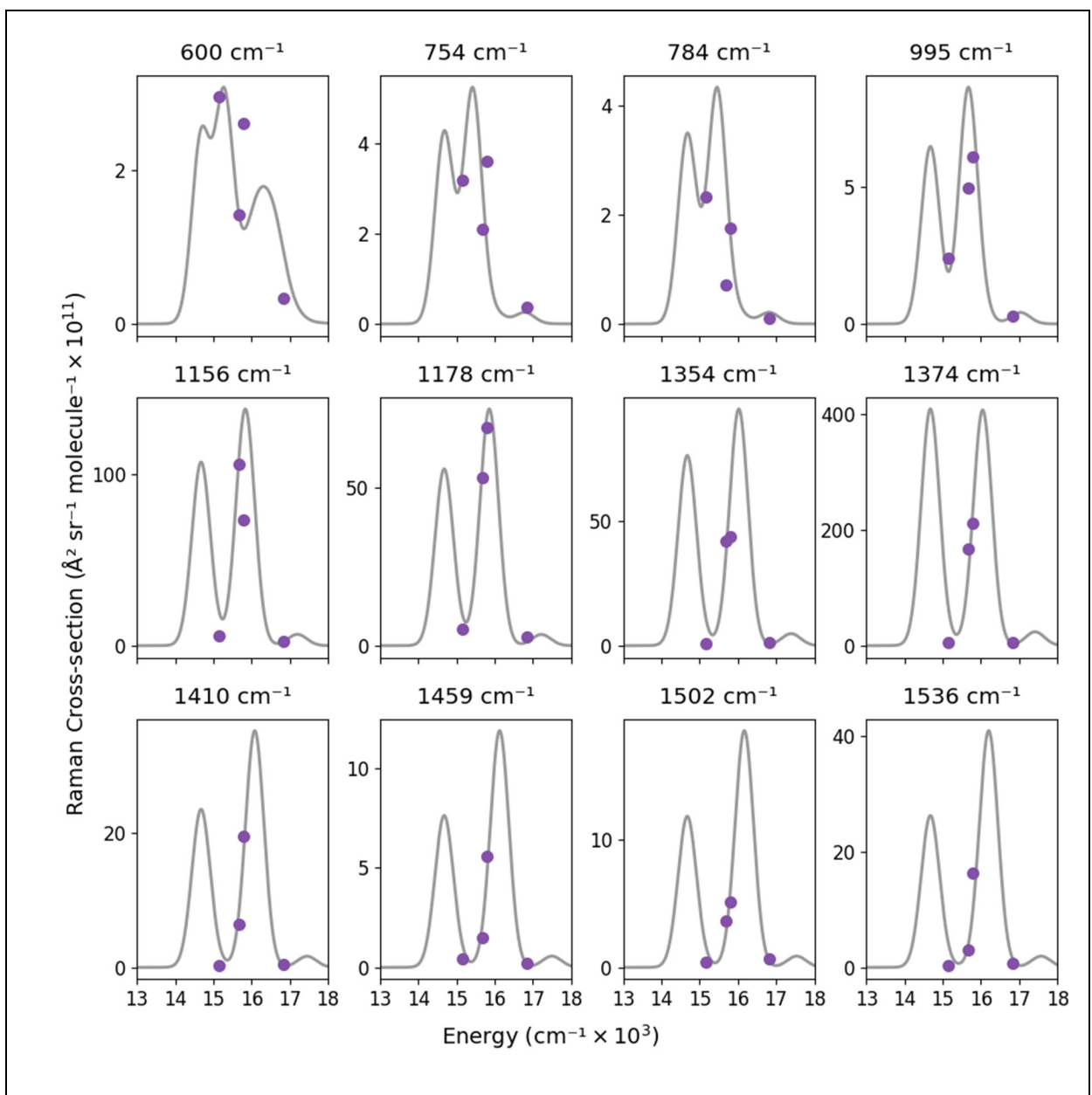

**Fig. S25.** Modeled (grey lines) and experimental (purple points) Raman excitation profiles for all Raman active vibrational modes of the 39 nm pentacene cavity at 55°.

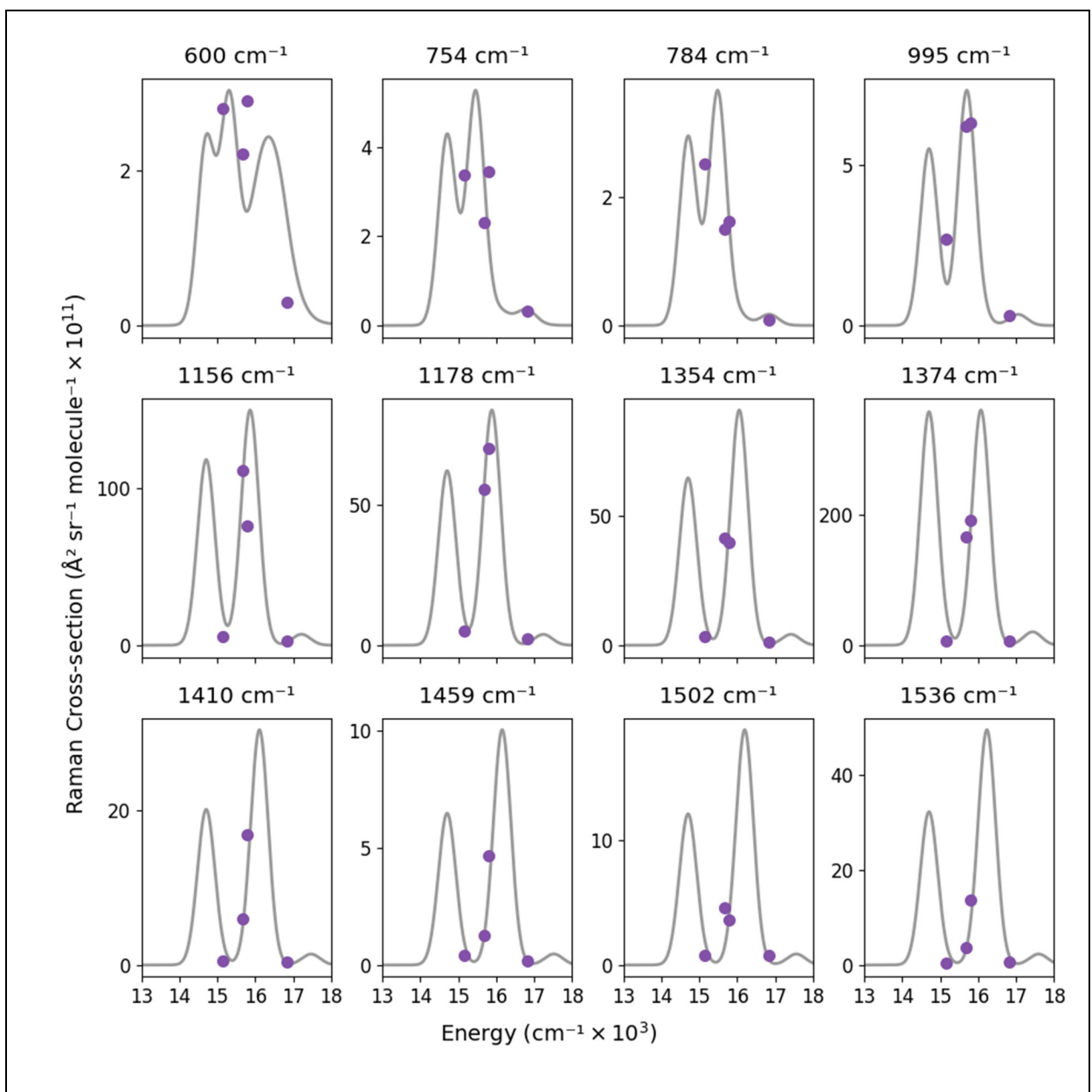

**Fig. S26.** Modeled (grey lines) and experimental (purple points) Raman excitation profiles for all Raman active vibrational modes of the 39 nm pentacene cavity at 60°.

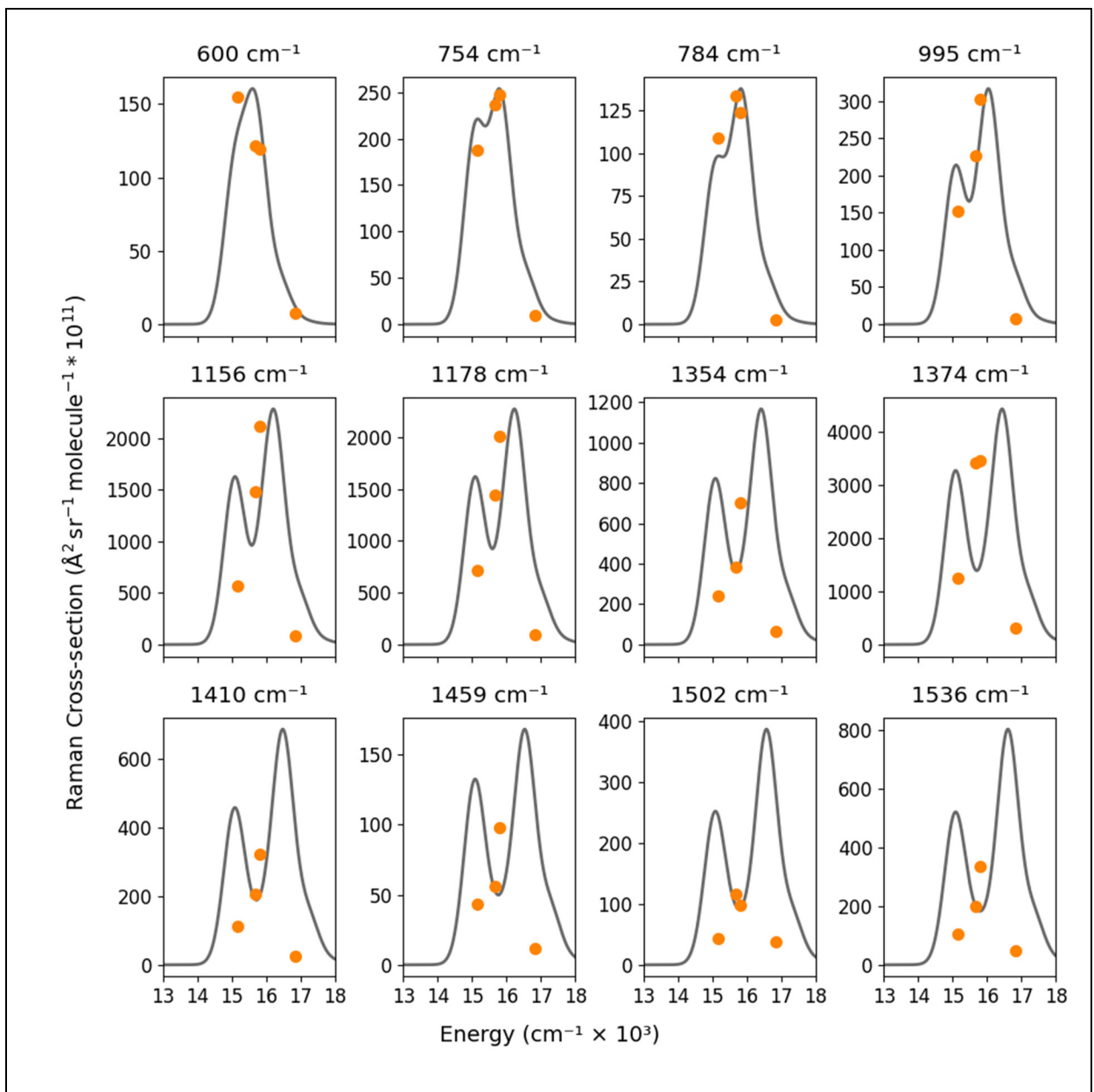

**Fig. S27.** Modeled (grey lines) and experimental (orange points) Raman excitation profiles for all Raman active vibrational modes of the 27 nm pentacene thin film at 30°.

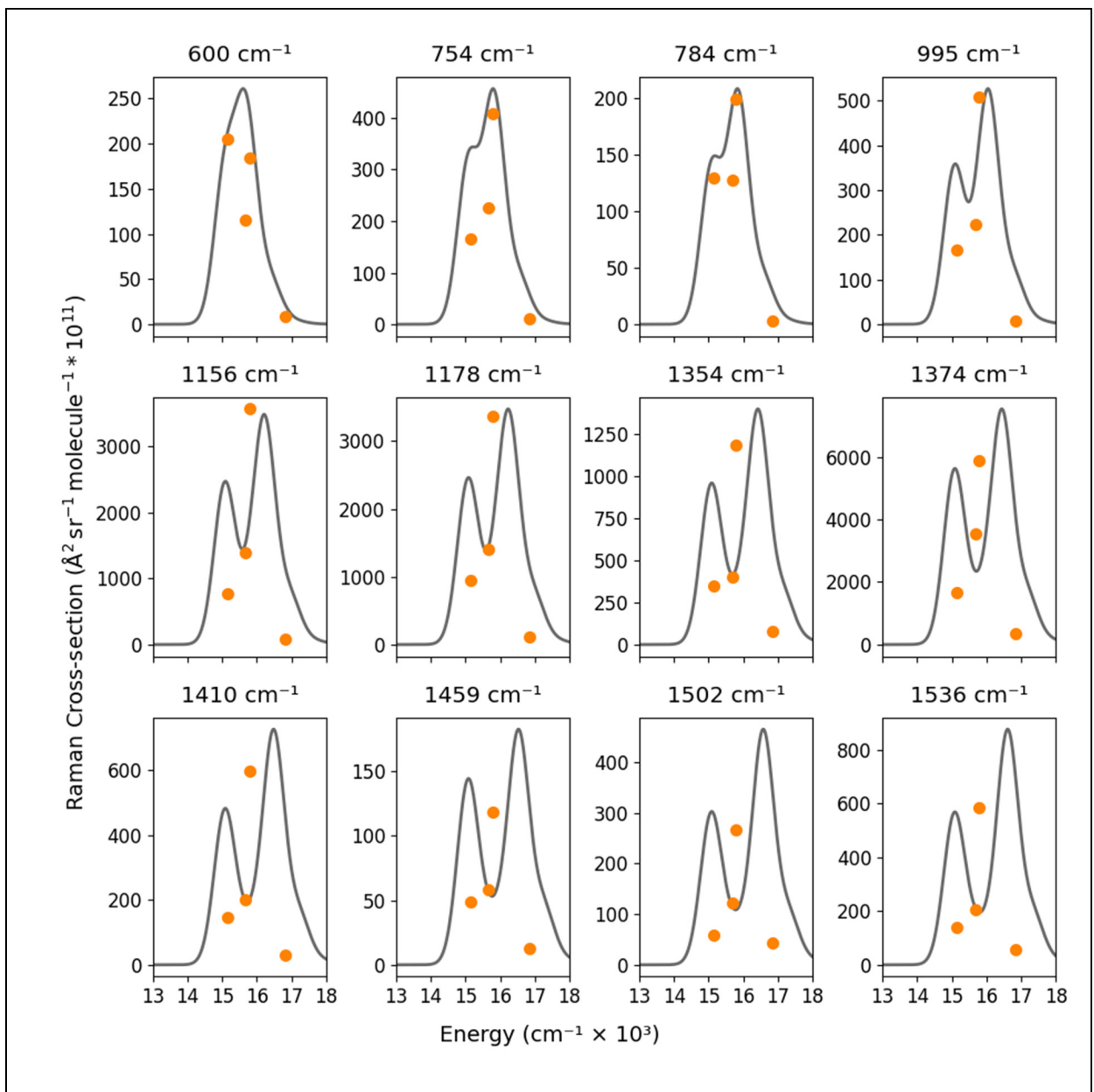

**Fig. S28.** Modeled (grey lines) and experimental (orange points) Raman excitation profiles for all Raman active vibrational modes of the 27 nm pentacene thin film at 45°.

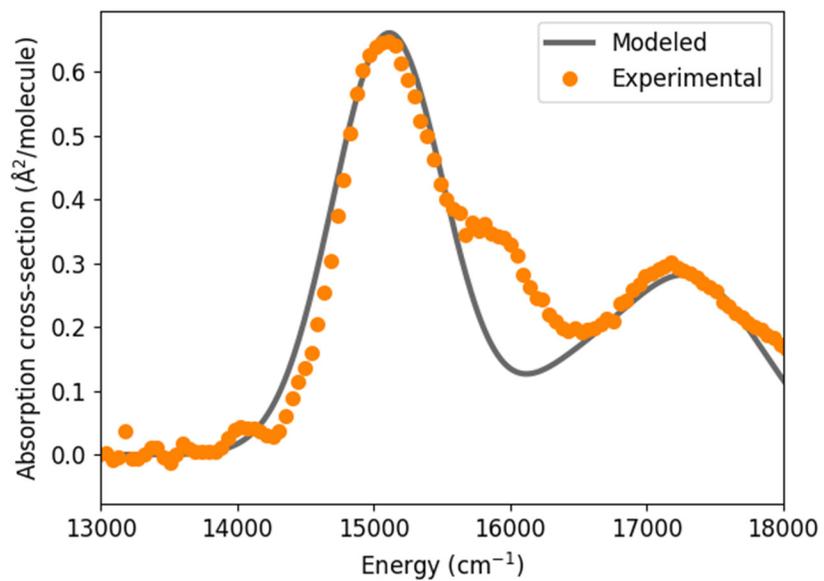

**Fig. S29.** Modeled (grey lines) and experimental (orange points) absorption spectra for the 27 nm pentacene thin film at 15° without the addition of the highly displaced frequency (730 cm$^{-1}$) in the RRIA analysis.

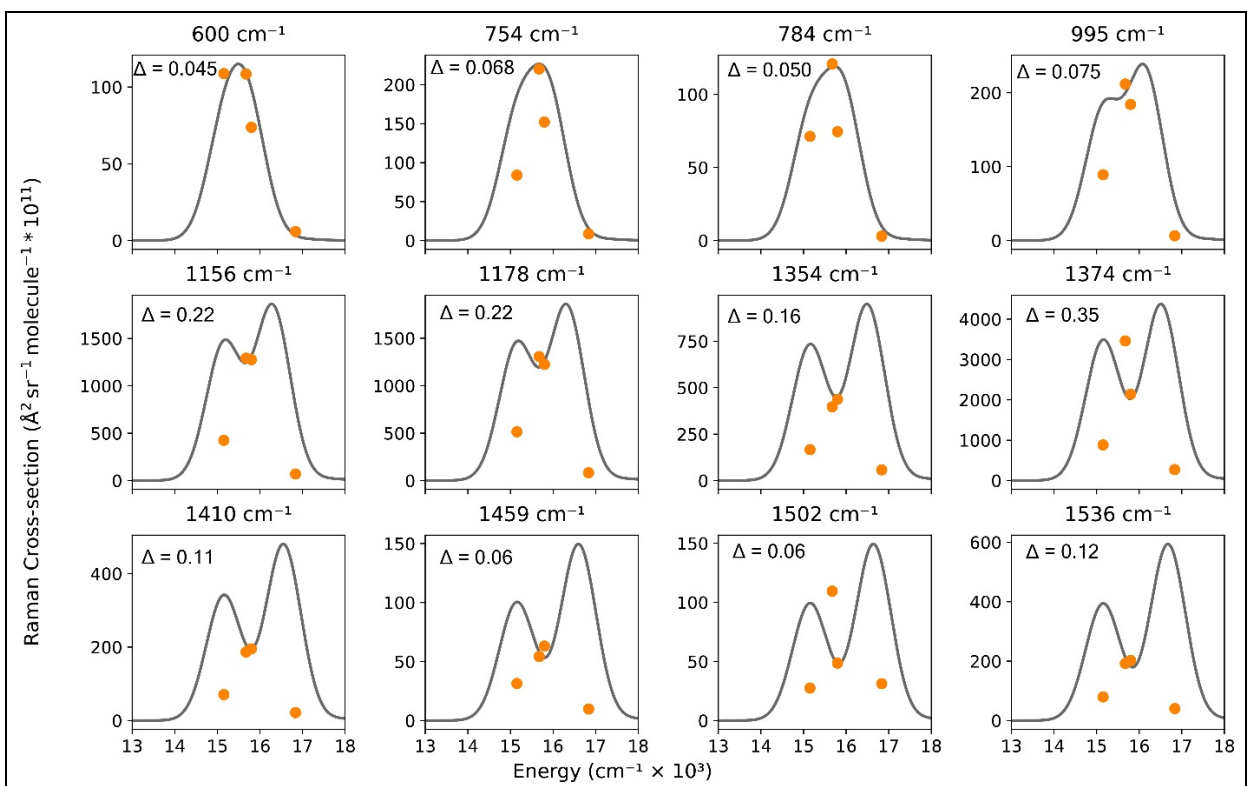

**Fig. S30.** Modeled (grey lines) and experimental (orange points) Raman excitation profiles for all Raman active vibrational modes of the 27 nm pentacene thin film at 15° without the addition of the highly displaced frequency (730 cm$^{-1}$) in the RRIA analysis.

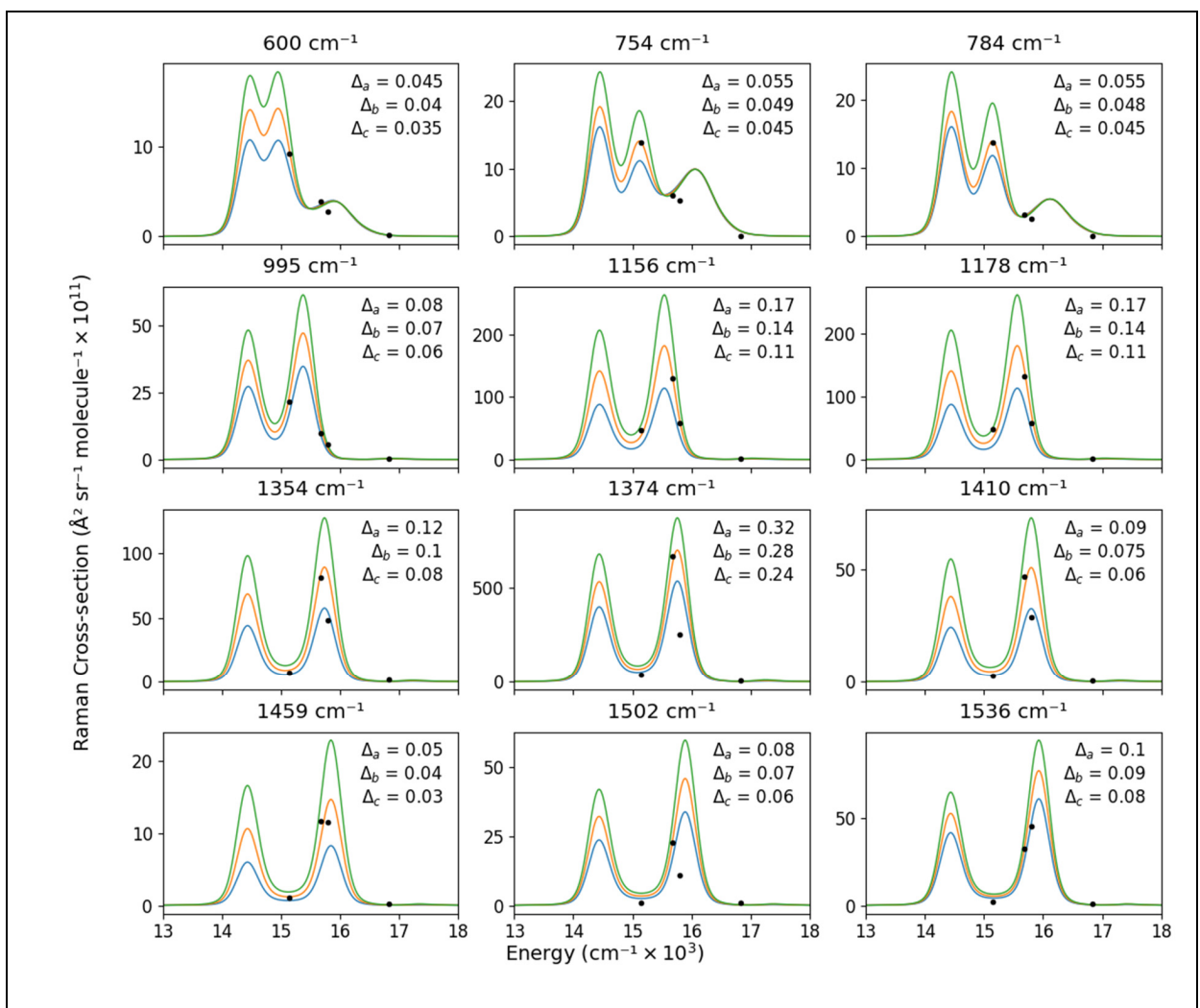

**Fig. S31.** Modeled (lines) and experimental (black points) Raman excitation profiles for all Raman active vibrational modes of the 27 nm pentacene cavity at 15° with green, orange, and blue curve modelled with $\Delta_a$, $\Delta_b$, and $\Delta_c$ values respectively. This plot highlights how the Raman excitation profiles are sensitive to the $\Delta$ values and the errors associated with the $\Delta$ values in our analysis.

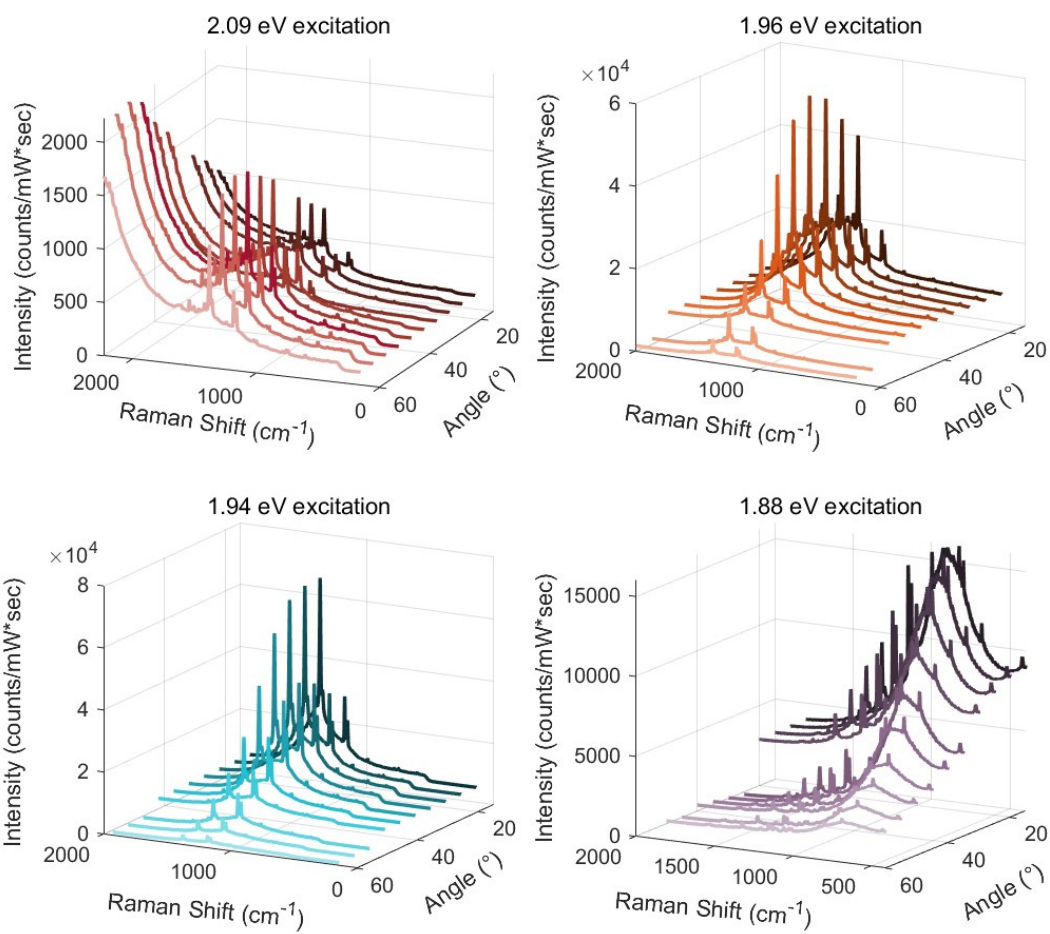

**Fig. S32.** Resonance Raman spectra of a 27 nm pentacene cavity measured at 15° to 60° in 5° increments for 2.09, 1.96, 1.94 and 1.88 eV excitation energies. To enhance visibility, the following spectra were scaled: 55° and 60° for 1.96 eV (factor of 2), 50°, 55°, and 60° for 1.94 eV (factor of 2), and 35°, 40°, 45°, 50°, 55°, and 60° for 1.88 eV (factors of 2, 2.5, 3, 3.5, 5, and 6, respectively).

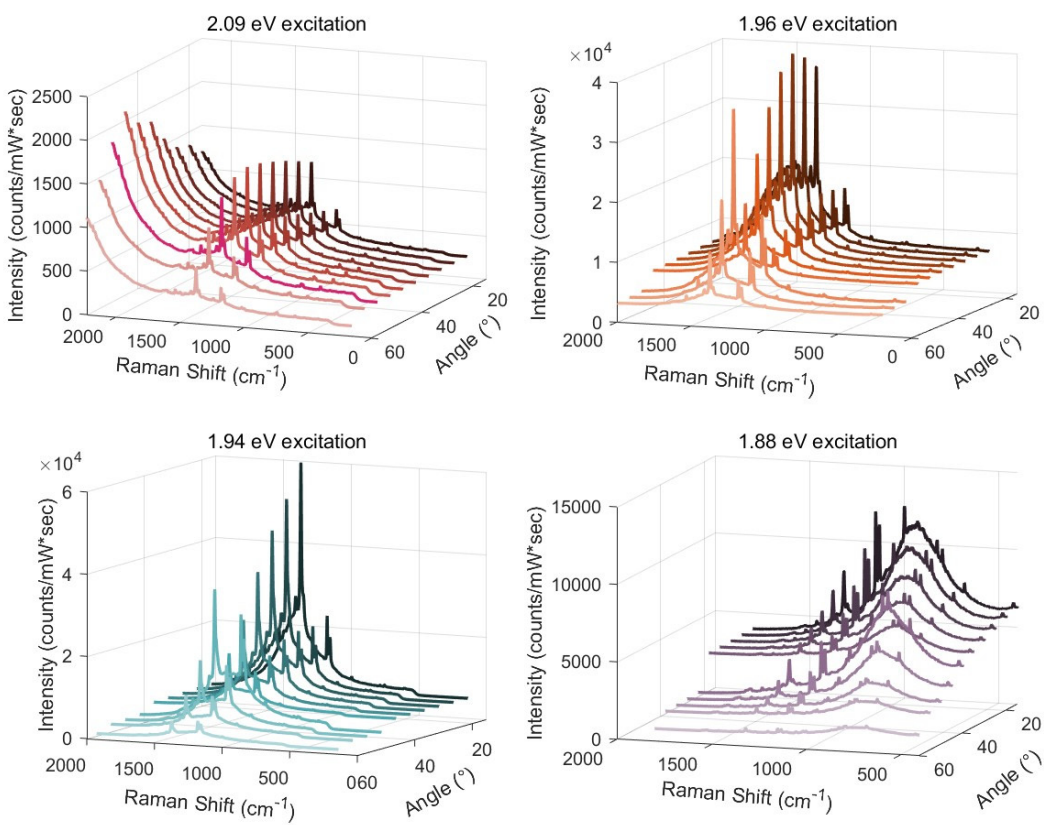

**Fig. S33.** Resonance Raman spectra of a 39 nm pentacene cavity measured at 15° to 60° in 5° increments for 2.09, 1.96, 1.94 and 1.88 eV excitation energies. To enhance visibility, the following spectra were scaled: 50°, 55°, and 60° for 1.96 eV (factor of 5), 45°, 50°, 55°, and 60° for 1.94 eV (factor of 5), and 40°, 45°, 50°, 55°, and 60° for 1.88 eV (factors of 4).

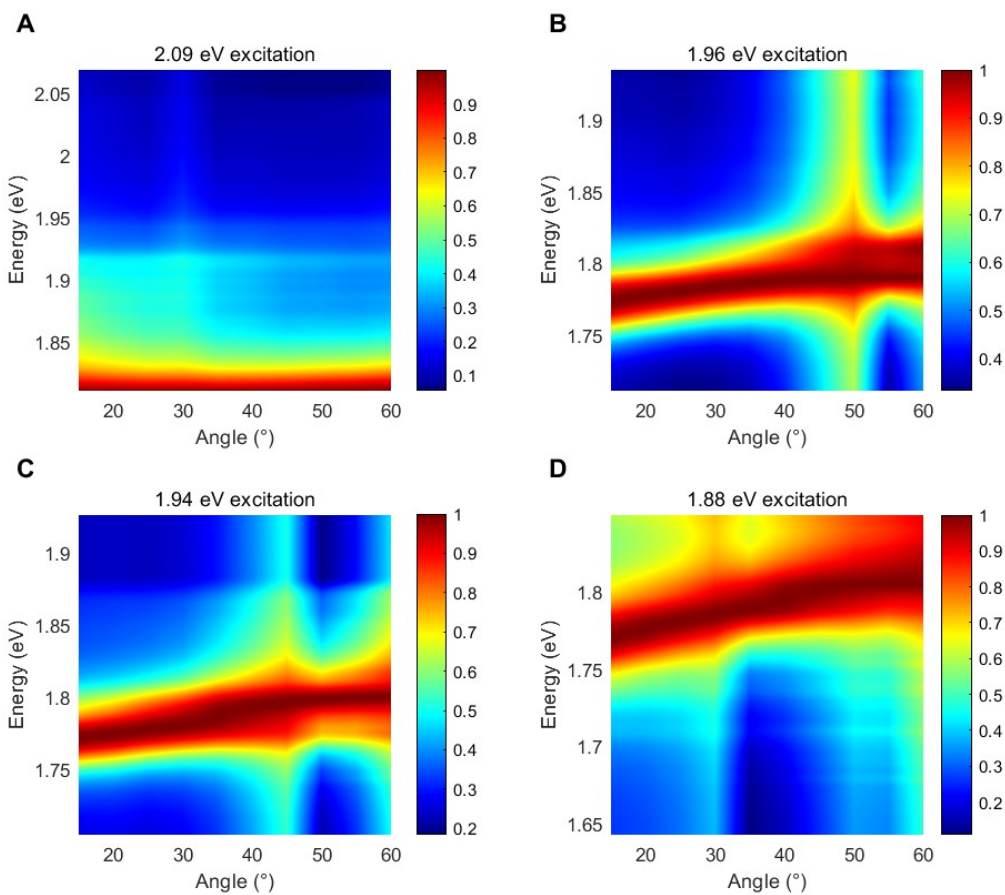

**Fig. S34.** Normalized Photoluminescence (PL) spectra collected from a cavity containing 27 nm of pentacene. Spectra were obtained by subtracting Raman background from measurements acquired using excitation wavelengths of **(A)** 2.09 eV, **(B)** 1.96 eV, **(C)** 1.94 eV, and **(D)** 1.88 eV.

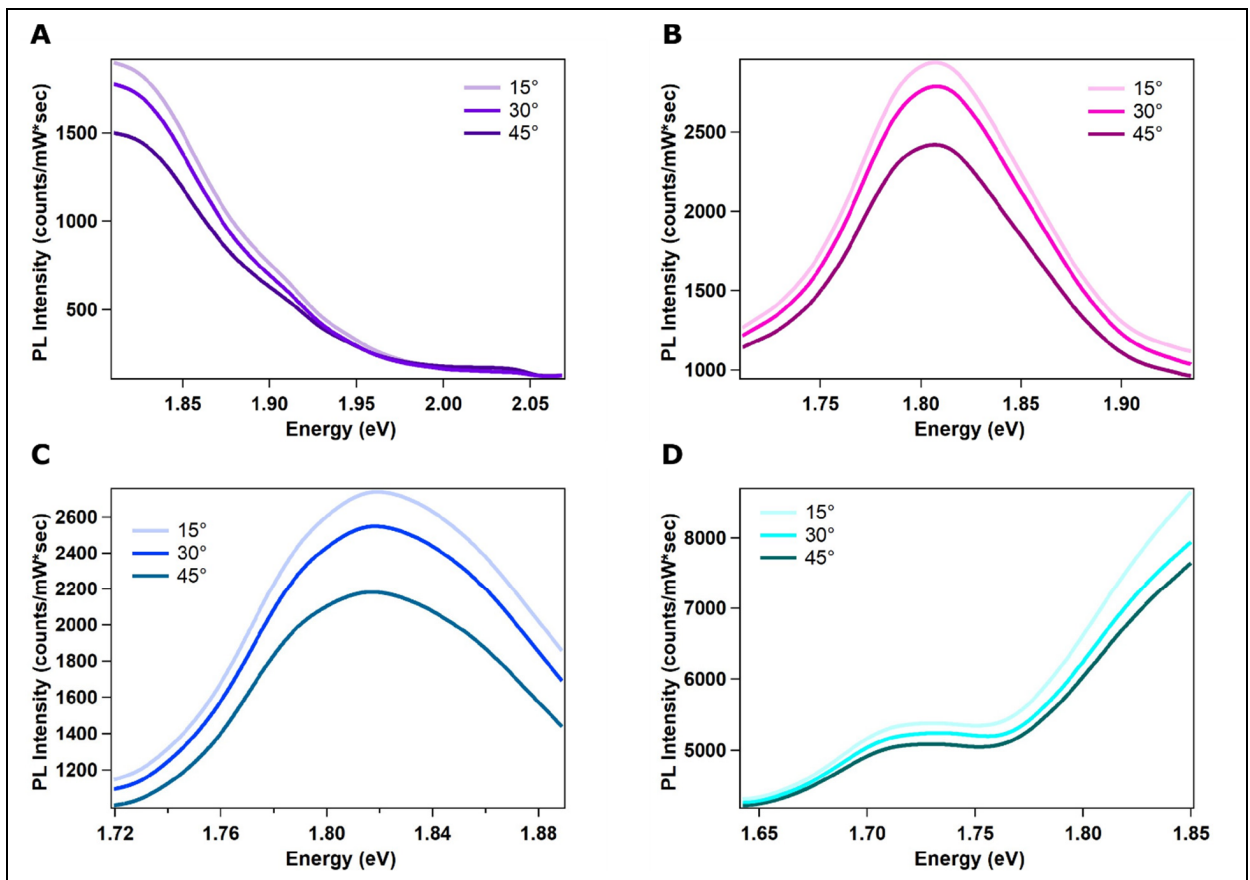

**Fig. S35.** Photoluminescence (PL) spectra of a 27 nm pentacene thin film. Spectra were obtained by subtracting Raman background from measurements acquired using excitation wavelengths of **(A)** 2.09 eV, **(B)** 1.96 eV, **(C)** 1.94 eV, and **(D)** 1.88 eV.